\documentclass[a4paper, 12pt]{article} 
\usepackage[right=1in,left=1in,top=1in,bottom=1in]{geometry}
\usepackage{graphicx} 
\usepackage{braket}
\usepackage{mathrsfs}
\usepackage[T1]{fontenc} 
\usepackage{amsmath}
\usepackage{pxfonts}
\usepackage[T1]{fontenc}
\usepackage{amssymb}
\usepackage{natbib}
\usepackage{epstopdf}
\usepackage{setspace}
\usepackage{enumitem}
\usepackage{sectsty}
\usepackage{wrapfig} 
\sectionfont{\large}
\bibpunct{(}{)}{,}{a}{}{,}
\usepackage{tikz}   
\usepackage{tikz-3dplot} 

\usepackage[protrusion=true,expansion=true]{microtype} 

\newtheorem{theorem}{Theorem}[section]

\newcommand{\qed}{\nobreak \ifvmode \relax \else
      \ifdim\lastskip<1.5em \hskip-\lastskip
      \hskip1.5em plus0em minus0.5em \fi \nobreak
      \vrule height0.75em width0.5em depth0.25em\fi}

\usepackage{bbm}
\usepackage{MnSymbol}
\usepackage[all]{xy}

\usepackage{xcolor,colortbl,array,amssymb}

\usepackage{epigraph}

\makeatletter
\newcommand{\x}[1]{{#1}}

\usepackage{tcolorbox}

\title{\textbf{Time's Arrow and Self-Locating Probability}} 

\author{Eddy Keming Chen\thanks{Department of Philosophy,  University of California, San Diego, 9500 Gilman Dr, La Jolla, CA 92093-0119. Website: www.eddykemingchen.net. Email: eddykemingchen@ucsd.edu  }}

\date{\textit{Philosophy and Phenomenological Research} } 

\begin{document}
\bibliographystyle{plain}

\maketitle 

 \epigraph{We therefore conclude that the universe is not a fluctuation, and that the order is a memory of conditions when things started. This is not to say that we understand the logic of it.}{Richard Feynman (1963)}



\begin{abstract}

One of the most difficult problems in the foundations of physics is what gives rise to the arrow of time.  Since the fundamental dynamical laws of physics are (essentially) symmetric in time,  the explanation for time's arrow must come from elsewhere. A promising explanation introduces a special cosmological initial condition, now called the \emph{Past Hypothesis}: the universe started in a low-entropy state. Unfortunately, in a universe where there are many copies of us (in the distant ``past'' or the distant ``future''), the Past Hypothesis is not enough; we also need to postulate self-locating (\textit{de se}) probabilities. However, I show that we can similarly use self-locating probabilities to strengthen its rival---the \emph{Fluctuation Hypothesis}, leading to in-principle empirical underdetermination and radical epistemological skepticism. The underdetermination is robust in the sense that it is not resolved by the usual appeal to `empirical coherence' or `simplicity.' That is a serious problem for the vision of providing a completely scientific explanation of time's arrow.
\end{abstract}

\hspace*{3,6mm}\textit{Keywords:  Past Hypothesis, the Mentaculus, Principle of Indifference,  Principal Principle, self-locating probabilities, de se, time's arrow, entropy, Bayesian epistemology, confirmation, empirical underdetermination, skepticism}   

\newpage

\begingroup
\singlespacing
\tableofcontents
\endgroup

\vspace{10pt} 







\nocite{ sep-time-thermo, albert2000time, loewer2016mentaculus, albert2015after, carroll2010eternity, pettigrew2014deference, lebowitz2008time, goldstein2001boltzmann, ismael2008raid, penrose1999emperor, price1997cosmology, winsberg2012bumps, pettigrew2016accuracy, fitelson2006does, pettigrew2014accuracy, loewer2004david, north2010empirical, fitelson2007likelihoodism, feynman2011feynman, coen2010serious, loewer2001determinism, ninan2010se, price1997time}

\section{Introduction}

One of the most difficult problems in the foundations of physics is what gives rise to the arrow of time.\footnote{In this paper, I use the phrase ``the arrow of time'' to refer to the entropic asymmetry of time that the entropy of the universe is lower in the past and higher in the future. It may be related to the ``flow of time'' or the ``passage of time.'' Precisely how they are related is outside the scope of this paper.  } 
On the one hand, nature seems to display a striking pattern of temporal asymmetry. An ice cube in a cup of hot coffee will melt; gas molecules contracted to a corner of the room will spread out; and a banana on the kitchen table will turn black. These processes (and many others) have a preferred temporal direction: they only happen from the past to the future. We do not see them happen in the other direction: an ice cube does not spontaneously form in a cup of hot coffee; gas molecules do not spontaneously contract to a corner; and a banana does not become fresher after a week. 



Physical states of the unmelted ice cube, the contracted gas molecules, and the fresh banana are less ``disorderly'' than those of the melted ice cube, the dispersed gas molecules, and the decayed banana. The former states have less \emph{thermodynamic entropy} than the latter states. The entropic arrow of time is defined as the direction of entropy increase in time. Such an entropy increase is summarized in the Second Law of Thermodynamics:
\begin{description}
\item[The Second Law] The (thermodynamic) entropy of a closed system (typically) does not decrease over time. 
\end{description}
On the other hand,  most candidates of  fundamental dynamical laws of physics are symmetric in time. Consider for example $F=ma$ with Newtonian gravitation. For any sequence of particle configurations that obeys $F=ma$, the time-reversal of that sequence also obeys $F=ma$: one simply needs to reverse the direction of the final particle velocities to get back to the initial state (with the opposite velocities). Similarly, the Sch\"odinger equation of quantum mechanics and the Einstein equation of general relativity are also symmetric in time (although their time-reversal operations are somewhat different). The fundamental dynamical laws of physics allow an ice cube to melt and also to spontaneously form in a cup of coffee. They are  insensitive to the past/future distinction found in macroscopic systems. 

Therefore, the arrow of time cannot come from the fundamental dynamical laws alone. Indeed, the standard explanation of time's arrow makes use of a special initial condition. It is a plausible idea: if we start a physical system in a very low-entropy state (unmelted ice, contracted gas, and fresh bananas),  the dynamical laws will (almost surely) take it to a higher-entropy state.\footnote{The qualifier ``almost surely'' is discussed in \S2.1.} However, it is rather complicated to postulate a low-entropy initial condition for every physical system. Instead, we can postulate a low-entropy initial condition for the whole universe, which  (we might accept on the basis of some plausibility arguments and some rigorous mathematical proofs) will likely lead to an increase of entropy for the whole universe as well as an increase of entropy for typical subsystems in the universe. This  low-entropy initial condition for the universe is now called the \emph{Past Hypothesis} (Albert 2000). 

The explanation of time's arrow in terms of the Past Hypothesis is suggested by \cite{boltzmann2012lectures}[1898]\S89 (though he ultimately seems to favor what may be called the \textit{Fluctuation Hypothesis}) and has many advocates:  \cite{feynman2011feynman}[1963], \cite{feynman2017character}[1965],  \cite{lebowitz2008time}, \cite{penrose1979singularities}, \cite{albert2000time},  \cite{callender2004measures}, \cite{north2011time},  \cite{wallace2011logic},  \cite{loewer2016earlymentaculus, loewer2016mentaculus}, \cite{goldstein2019gibbs}, and myself \citep{chen2020harvard, chen2018NV}.\footnote{ See \cite{earman2006past} for some worries about  the Past Hypothesis as a hypothesis about the initial condition for the universe. See \cite{goldstein2016hypothesis} for a discussion about the possibility and some recent examples, of explaining the arrow of time without the Past Hypothesis.} We all agree that some version of the Past Hypothesis (PH) should be postulated in the fundamental physical theory. Setting aside future progress in cosmology,\footnote{See, for example, the interesting ideas of \cite{carroll2004spontaneous}. At the moment their proposal is quite speculative, but it is conceptually illuminating as a possible alternative to the PH paradigm shared by the previously quoted people. See \cite{goldstein2016hypothesis}  for some discussions of the ideas of Carroll and Chen and those of Julian Barbour. } the PH explanation is a promising one for understanding time's arrow in our universe. Moreover, it has been argued that the explanation can also justify our standard inferences to the past.  Thus, the scientific explanation may  extend to the epistemic realm:  our ordinary beliefs about the past obtain their justification partly in virtue of PH.\footnote{Feynman has a beautiful way of describing this, in the text right after the epigraph \cite[46-5]{feynman2011feynman}: ``For some reason, the universe at one time had a very low entropy for its energy content, and since then the entropy has increased. So that is the way toward the future. That is the origin of all irreversibility, that is what makes the processes of growth and decay, that makes us remember the past and not the future, remember the things which are closer to that moment in the history of the universe when the order was higher than now, and why we are not able to remember things where the disorder is higher than now, which we call the future.''} Based on these reasons, \cite{albert2000time, callender2004measures} and \cite{loewer2016earlymentaculus, loewer2016mentaculus}  suggest that PH (and an accompanying statistical postulate) should be understood as a candidate fundamental law of nature.

The Past-Hypothesis explanation would be immensely powerful if it were successful. One of its attractions, as suggested by Albert and Loewer, is its realization of a bold and ``fundamentalist'' vision that  time's arrow is to be scientifically explained in terms of objective laws of fundamental physics (or objective postulates of fundamental physics), and such laws (or postulates) can be empirically confirmed by our current evidence. The explanation is supposed to be entirely empirical. It seems superior to a metaphysical one that postulates a fundamental arrow of time and justifies it on a priori reasons. As such, the success of the PH explanation would be philosophically significant. 

 Some philosophers, such as \cite{winsberg2004can} and \cite{earman2006past}, have raised \textit{technical} objections against the explanation. In contrast, I focus on two \textit{conceptual} worries about self-locating (\textit{de se}) probabilities that are  troublesome even for those of us who defend the PH research program and remain optimistic that the technical objections can ultimately be overcome. Resolving the conceptual worries requires deeper engagement among epistemologists and philosophers of physics.  First, PH (and the accompanying statistical postulate) is not enough. We  need to postulate certain self-locating (\textit{de se}) probabilities, by using a temporally biased self-locating probability distribution.  That already is problematic for the original vision, since it is unclear whether self-locating probabilities can be part of the objective physical laws.  Second and more surprisingly,  if we allow self-locating probabilities to appear in the physical theory, we open a Pandora's box:  they lead to empirical underdetermination and epistemological skepticism. That calls into question the ambitious vision of providing a completely scientific explanation of time's arrow. 

 First (in \S 2), I argue for the Self-Location Thesis: the PH-explanation of time's arrow requires a postulate about self-locating probabilities.\footnote{ Similar ideas are  considered by  \cite{winsberg2010, winsberg2012bumps}, but his focus and arguments are different from mine in the following ways. In his 2010, he suggests that the PH-explanation of time's arrow requires a self-locating proposition that we are located  between the time of PH and the first thermodynamic equilibrium. He argues that the necessity of postulating this self-locating proposition is a problem for Loewer's thesis that the \textit{lawfulness} of special sciences (such as biology, psychology, and economics) is grounded in the fundamental laws. Winsberg ultimately argues in favor of the autonomy of the special sciences. I do not focus on the special sciences in this paper. Moreover, he does not draw the connection to self-locating probabilities as I do here. In his 2012, he uses the self-locating proposition in his cost-benefit analysis of a different explanation of time's arrow---Carroll-Chen's time-symmetric model (\citeyear{carroll2004spontaneous}) in which a ``mother universe'' that has no  entropy maximum constantly gives birth to ``baby universes'' starting in sufficiently low entropy. I do not discuss the Carroll-Chen model except to set it aside.  }  (Here and elsewhere in the paper, ``self-locating'' and ``de se'' are used interchangeably.) 
 In a universe that persists long enough, there may be many copies of us (e.g. produced by random fluctuations).  For reasons similar to the Boltzmann Brain argument (if we are equally likely to be any observers with our mental states, then it is most likely we are observers produced by the smallest fluctuations---the Boltzmann brains), we need to postulate that our current time is located between the Big Bang and the first equilibrium (with uniform probabilities over any observers that share our features). 
This part may be familiar to experts in statistical mechanics, but I intentionally explain it slowly for two reasons: (1) some readers with philosophical expertise may not be familiar with the technical issues, and (2)  we need to be as clear as we can to fully understand the philosophical issues we face. The next point builds on the first one. 

Second (in \S 3), I argue that, if it is permissible to save the Past-Hypothesis explanation by adding self-locating probabilities, we can do the same to strengthen a rival--the Fluctuation Hypothesis--and achieve in-principle empirical underdetermination by our current evidence.   In retrospect this move may seem  obvious, but as far as I know it is new. A surprising consequence is that their empirical underdetermination  leads to radical skepticism. The strengthened version of the Fluctuation Hypothesis resembles the infamous Boltzmann Brain Hypothesis. However, I explain that the augmented version with what I call ``Boltzmann bubbles'' is more believable and epistemically conservative. It  can even accommodate various epistemological positions that are ruled out by the Boltzmann Brain Hypothesis. 

The paper can be seen as presenting a dilemma. Either self-locating probabilities have a place in a scientific explanation or they do not. On my view,  both horns can be problematic for the original vision. If they do not have a place in a scientific explanation,  the PH-explanation is insufficient. If they do have a place, there is in-principle empirical underdetermination such that two equally simple (or complex) scientific theories say contradictory things about time's arrow. Thus, postulating self-locating probabilities to explain time's arrow turns out to be doubly problematic. My conclusions are conditional on the assumptions, and the particular skeptical conclusion might be avoided by appealing to certain traditional epistemological responses. What response to choose requires careful analysis of the case. Addressing these worries in a thorough way may require expertise in epistemology and general philosophy of science. Where possible, in the paper I try to explain the concepts ``from the ground up.'' I hope the present analysis will lead to further work on this issue by epistemologists and  philosophers of science. 

In this paper, I assume that the universe allows many thermodynamic fluctuations. It is an open question whether there are such fluctuations in our universe, depending on whether the universal state space is finite or infinite (in terms of phase space volume or  Hilbert space dimension). Even if the assumption were false of the actual universe,  it would still be worth investigating how serious the problem would be, and what strategies we would need, if Nature were not so kind to us. If the dilemma is worrisome enough, that would provide additional motivation to look for a theory where fluctuations are not as prevalent. 


\section{The Self-Location Thesis}

In this section, I  first review the standard explanation for time's arrow in terms of PH. I  introduce some concepts from philosophy of physics that may be unfamiliar to non-specialists. I  then consider whether it is the best theory that explains the evidence. To do so I  consider an alternative explanation without PH. Thinking about their differences leads us to consider the Self-Location Thesis. 

\subsection{The Mentaculus Theory}

According to the standard picture, PH is  key to understanding the apparent temporal asymmetry: from ice melting and gas dispersing to the more general statement about entropy increase in the Second Law of Thermodynamics. However, it is to be supplemented by two more postulates: the dynamical laws of physics (such as $F=ma$) and a probabilistic postulate called the \emph{Statistical Postulate}. Together, they provide the standard (probabilistic) explanation of time's arrow, i.e., the (typical) unidirectional change of entropy. 

To appreciate the explanation that PH provides, let us introduce some technical terms from statistical mechanics. PH ensures that the universe started in a  low-entropy initial condition.\footnote{In this paper I assume that the universe has a temporal boundary or a space-time singularity that can be called ``the beginning.'' To relax this assumption, as some cosmological theories does, will take us to more  issues than we have the space to discuss here.  } But what is entropy? Entropy is a macroscopic quantity that is on par with density, temperature, and pressure. It can be calculated by measuring the transformations in thermodynamic quantities.\footnote{For example, \cite{clausius1867mechanical} defines the change in entropy for an isolated system to be equal to the change in heat divided by temperature.} However, it can also be defined by making a distinction between  macrostates and  microstates. To illustrate, let us consider a classical-mechanical system of gas  in a box.\footnote{Below we will use some concepts from classical statistical mechanics. For the technically inclined, we note that the classical framework can be adapted to a quantum framework as follows:  use a Hilbert space instead of phase space, a state vector instead of a phase point, a subspace instead of a subset, and dimension counting instead of volume measure to measure entropy. See  \cite{goldstein2010approachB} for an informative overview of Boltzmannian quantum statistical mechanics. } A macrostate is a characterization of the gas in a box in terms of macroscopic variables such as pressure, volume, and temperature, while a microstate is a characterization of the gas in terms of microscopic variables such as the positions and velocities of all of the gas molecules. A macrostate is compatible with many possible microstates---a macrostate is multiply-realizable by many different microstates. The actual microstate realizes a particular macrostate, but it shares the macrostate with many other microstates that are macroscopically similar.

Intuitively, there are more microstates realizing the macrostate in which all the gas molecules are spread out uniformly than there are microstates realizing the macrostate in which all of them are contracted to a corner. 
According to Ludwig Boltzmann, (thermodynamic) entropy measures how ``many'' microstates are compatible with a given macrostate.\footnote{For the technically inclined, here are some formal details. The Boltzmann entropy of a microstate is proportional to the volume of the macrostate that it belongs to:
$$S_{B} (X) = k_B \text{log} |\Gamma_{M(X)}|,$$
where $X$ is the microstate of the system, $k_B$ is the Boltzmann constant, $\Gamma_{M(X)}$ is region of phase space that corresponds to the macrostate of $X$, and  $|\cdot |$ denotes the volume measure of the $6N$-dimensional phase space. It follows that the larger the macrostate, the higher the Boltzmann entropy.  If at $t_1$ the system is in $X_1$ which belongs to a small macrostate $M_1$, and at $t_2$ the system is in $X_2$ which belongs to a large macrostate $M_2$, then the Boltzmann entropy has increased from $t_1$ to $t_2$. The transition from $X_1$ at $t_1$ to $X_2$ at $t_2$ is determined by the classical laws of motion, i.e. Hamilton's equations which correspond to the familiar $F=ma$.} 
\begin{figure}
\center
\includegraphics[scale=0.65]{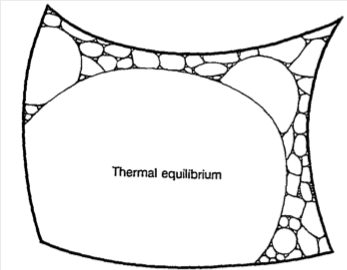}
\caption{This picture from Penrose (1989) is a schematic depiction of the phase space (restricted to the energy hypersurface). Each point corresponds to an exact microstate. Each bounded region corresponds to a macrostate, which is a set of macroscopically indistinguishable microstates. The equilibrium macrostate (of maximum entropy) takes up overwhelmingly most volume; all other macrostates are much smaller. The macrostates partition the phase space into regions. }
\end{figure}
It follows, on this definition of entropy, the ``spread-out'' state has higher entropy than the ``contracted'' state. The  uniformly spread-out state of gas in a box also corresponds to thermal equilibrium, the state of entropy maximum. We can plot all the microstates on a $6N-$dimensional space called the \emph{phase space}. The macrostates are sets of microstates that are macroscopically indistinguishable; they partition the space into distinct non-overlapping regions. The standard way of counting the microstates, since there is an infinity of them, is by using the standard Lebesgue measure on  phase space. On this way of counting,  the equilibrium state takes up the overwhelming majority of volume in phase space (see Figure 1).

In the language of  phase space, we can consider (as a first approximation) the entire universe to be  such a classical system. Then PH selects a special macrostate $M(0)$ to be where the initial microstate of the universe lies in. $M(0)$ is small in volume measure. Thus, it has very low entropy (by Boltzmann's definition). But it is still compatible with a continuous infinity of microstates. For a typical initial microstate lying in $M(0)$, it will follow the dynamical laws (such as $F=ma$) to evolve into other microstates. Since the overwhelming majority of microstates surrounding $M(0)$ will lie in macrostates of larger volume, typical trajectories from $M(0)$ will get into  larger macrostates, which corresponds to  an increase of entropy.\footnote{This is a place where rigorous results are difficult to obtain. The history of statistical mechanics contains many attempts to make progress in this direction. Boltzmann's not fully rigorous argument for his Boltzmann equation is one such step.  Oscar E. Lanford's celebrated  proof  of statistical results in a model of hard spheres and diluted gas  is another one. See \cite{uffink2010time} for more discussions and references. }

However, as Albert (2000) explains, PH is still not sufficient. Not all microstates lying in $M(0)$ will get into higher-entropy macrostates. Some of them are ``bad'': they will evolve under the dynamical laws into lower-entropy macrostates.\footnote{Their existence is suggested by the time-reversal invariance of the dynamical laws.}  We need a reason to neglect the bad microstates. The Statistical Postulate provides such a reason. It specifies a uniform probability distribution (with respect to  the volume measure) on  phase space. According to this probability distribution, the ``bad'' microstates receive very small weight. It is overwhelmingly likely that our world did not start in one of the ``bad'' microstates. Hence, it is with overwhelming likelihood that the entropy of our world has always been increasing in the past and will continue increasing in the future. (See Figure 2.)  This constitutes a probabilistic explanation of the entropic arrow of time. 

\begin{figure}
\center
\includegraphics[scale=0.45]{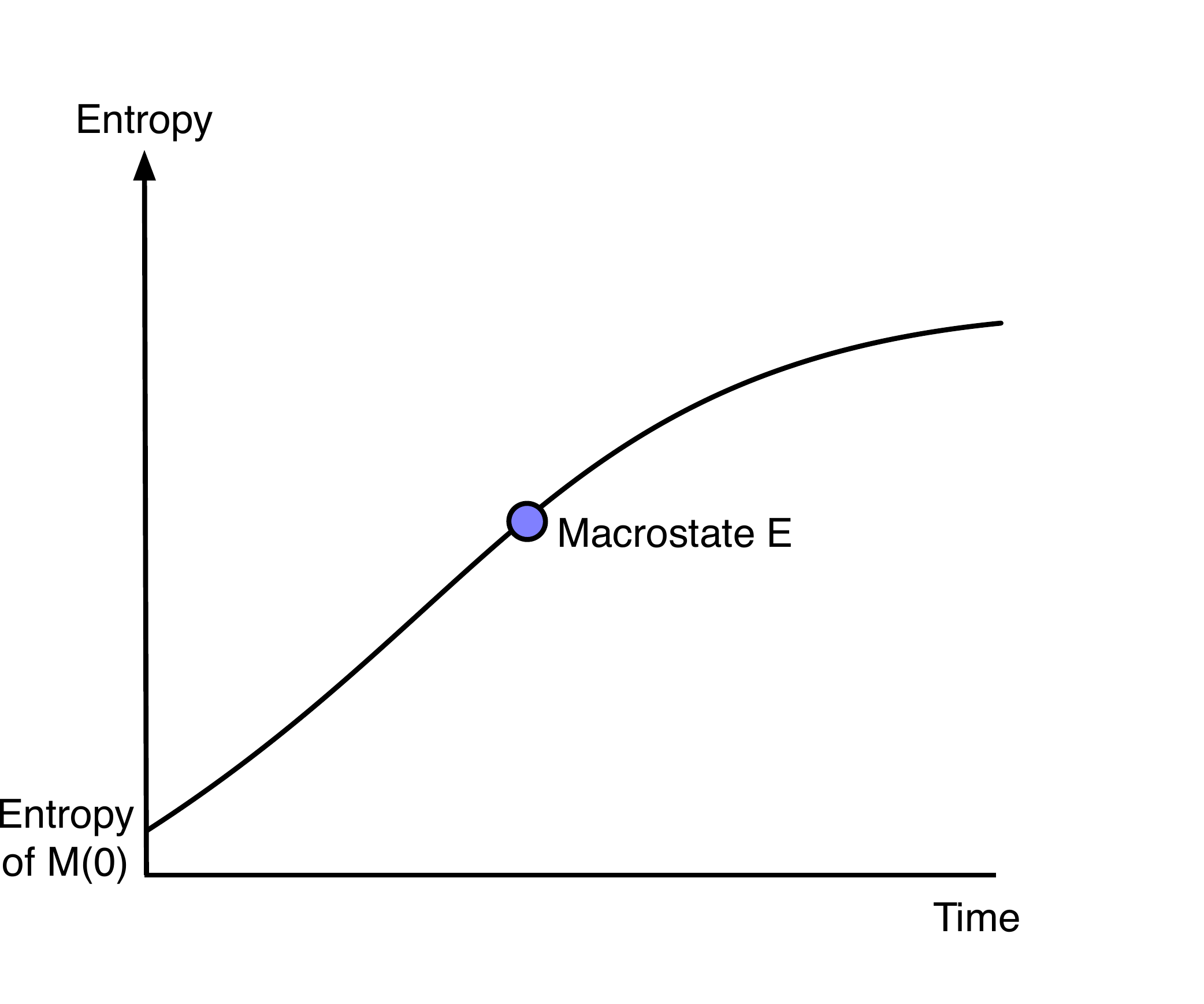}
\caption{The Past Hypothesis selects a low-entropy initial condition, a special macrostate $M(0)$. The macrostate $E$ stands for the macrostate of the universe at the time of observation. }
\end{figure}

Following Albert (2000) and Loewer (2016), let us call the following  the Mentaculus Theory ($T_M$)\footnote{The term ``Mentaculus Theory'' comes from the phrase ``Mentaculus Vision'' which is coined by Loewer (2016). It is based on the movie \emph{A Serious Man} (2009) directed by Ethan Coen and Joel Coen. In the movie, the title of Arthur's book is \emph{The Mentaculus}, which means the ``probability map of the universe.'' Here, the Mentaculus Vision is that, given the probability distribution of PH + SP, we can obtain a probability assignment of every proposition formulable in the language of  phase space or the Hilbert space. }: 
\begin{enumerate}
\item \textbf{Fundamental Dynamical Laws (FDL):} A specification of the evolution of the fundamental microstates of the universe (and the fundamental microstates of its isolated sub-systems). 
\item \textbf{The Past Hypothesis (PH)}: A specification of a boundary condition characterizing the universe's macrostate at the time of the Big Bang as M(0). In agreement with contemporary cosmology, M(0) is a macrostate with extremely low entropy. 
\item \textbf{The Statistical Postulate (SP)}: A uniform probability distribution (with respect to the volume measure) over the physically possible microstates that realize M(0).
\end{enumerate}
Together, these three postulates provide an explanation of time's arrow. PH and SP constitute an extremely biased initial probability distribution on  phase space. They make it overwhelmingly likely that the microstate of our universe lies on a trajectory that will go to higher-entropy states.

\subsection{The Fluctuation Theory}

The Mentaculus Theory ($T_M$)  explains the origin of temporal asymmetry  in terms of a specially chosen initial condition and a probability distribution---PH and SP. It has an ambitious goal of explaining a wide range of phenomena: ice melts in room temperature; things grow and decay; and many other temporally asymmetric phenomena (and perhaps even the asymmetry of records and control: that we have records of the past but not of the future, and we currently have  control of the future but not of the past\footnote{See \cite{albert2000time} and Loewer (2016) for arguments that connect the Mentaculus Theory to  these other arrows of time.}). This theory could be our best guide to time's arrow. 

What are our grounds for endorsing $T_M$? Defenders of this theory (such as Albert and Loewer) suggest that it is our best explanation for time's arrow. But can we  rule out all other possible explanations? Is PH really better than its alternatives?  The answers are not so simple. 

 Let us consider an alternative to $T_M$---the Fluctuation Hypothesis. This alternative is related to suggestive remarks made by Ludwig Boltzmann (1964 (1899), \S 90). Boltzmann might have endorsed it  at some point. The motivation is as follows. It seems that the special initial condition in $T_M$ is too special and too contrived. \cite{roger1989emperor} estimates that the initial macrostate $M(0)$ is tiny compared to the available volume on  phase space. A rough calculation based on classical general relativity suggests that the PH macrostate (specified using the Weyl curvature) is only $\frac{1}{10^{10^{123}}}$ of the total volume in phase space. It seems more satisfactory to explain why the universe started in such a special state than to postulate it as axiomatic as on $T_M$.

One way to explain PH is based on thermodynamic fluctuations. The universe, considered as a closed thermodynamic system, will typically increase in entropy and remain the same after it reaches the entropy maximum. However, sometimes it will  decrease in entropy, producing a thermodynamic fluctuation---a deviation from the normal behavior. Thermodynamic fluctuations are rare, but they do occur given enough time. This is the reason that the Second Law of Thermodynamics should be regarded as a probabilistic (or statistical) law that holds with overwhelming probability but not with certainty. 

An extreme kind of fluctuation is demonstrated by Poincar\'e to occur in some systems. The Poincar\'e Recurrence Theorem says, roughly, that if we start from anywhere in  phase space, we will (almost surely) come back to it infinitely many times.\footnote{For the mathematically inclined, here is a rigorous statement of the theorem. Let $(X, \mathscr{B}, \mu)$ be a measure space. Let $T: (X, \mathscr{B}) \rightarrow (X, \mathscr{B})$ be a function such that $\forall A \in \mathscr{B}, T^{-1}(A) \in \mathscr{B}$. Definition: the measure $\mu$ is a $T$-invariant measure if $\forall A \in \mathscr{B}, \mu(A)=\mu(T^{-1}(A)).$
\begin{theorem}
\emph{(The Poincar\'e Recurrence Theorem)}
Let $\mu$ be a $T$-invariant measure with $\mu(X) < \infty$. $\forall A \in \mathscr{B}$ such that $\mu(A)>0$, we have a.e. $x\in A$ such that $\# \{n\in \mathbb{Z}^{+} | \text{ } T^{n} (x) \in A \}=\infty.$
\end{theorem}
} 
The argument, however, assumes that the phase space of the system is bounded with finite volume. This assumption may or may not apply to our universe. It is still an open question whether there can be such dramatic kinds of fluctuations, i.e. recurrences, in our universe. The question depends on whether the universal state space (phase space or Hilbert space) is infinite. However, even if we cannot guarantee the existence of recurrences by something like the Poincar\'e Recurrence Theorem, we cannot rule out the existence of less dramatic fluctuations and more localized fluctuations. Recurrences are sufficient but not necessary for the Fluctuation Hypothesis, which only attempts to explain our observation by some random fluctuation (that can be much smaller than a full recurrence).  In any case, this is not the place to settle the technical question.\footnote{For some recent work, see \cite{carroll2017boltzmann} for discussions about the possibility of a finite-dimensional Hilbert space. } The goal of the paper is rather conceptual. As mentioned earlier, even if we were to have empirical grounds to entirely rule out the possibility of fluctuations in the actual universe, it would still be interesting to investigate how serious the problem would be if fluctuations were possible. 

\begin{figure}[t]
\center
\includegraphics[scale=0.45]{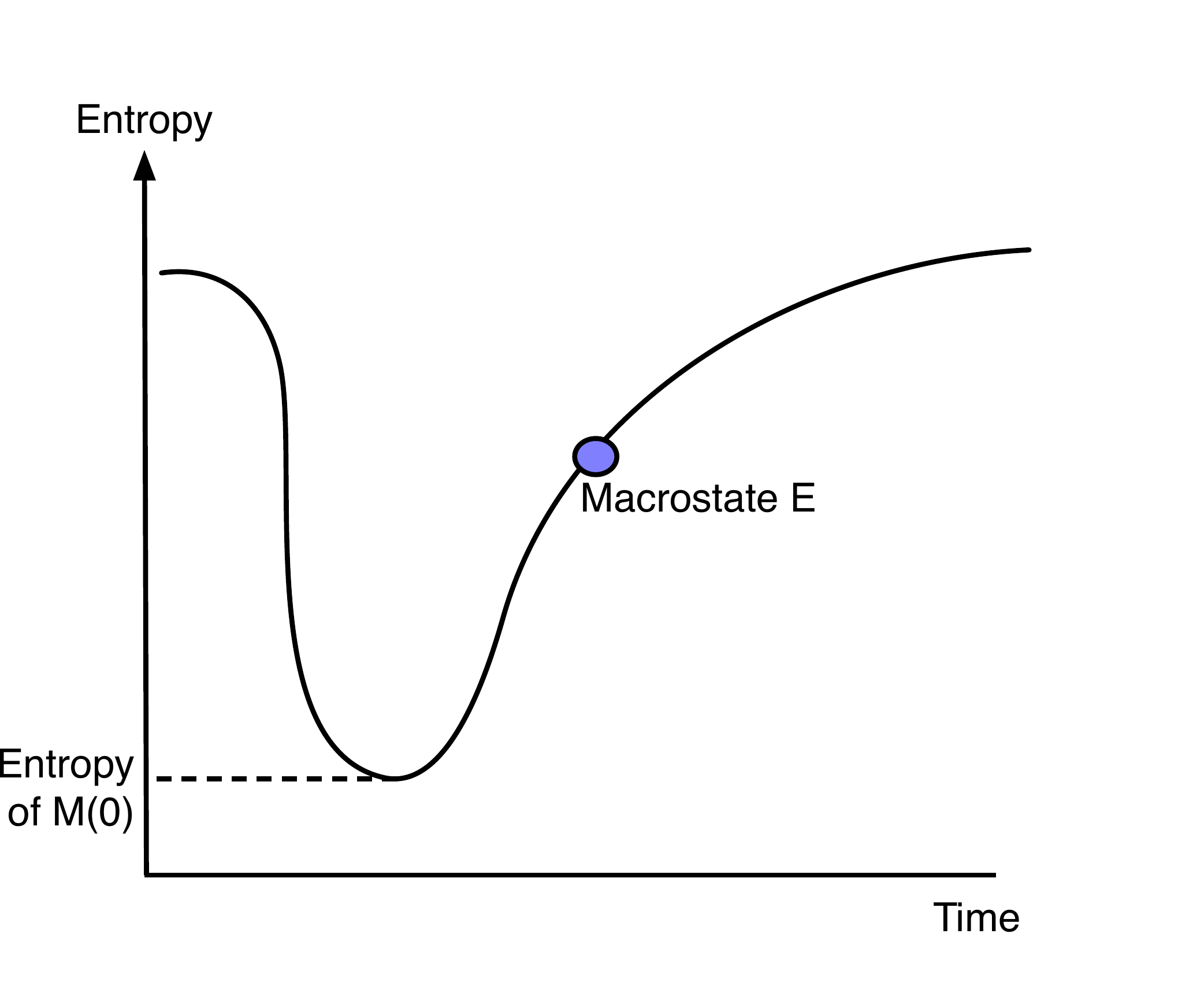}
\caption{The Fluctuation Hypothesis allows the universe to start in any macrostate. But the theory predicts that the universe will eventually fluctuate to a low-entropy macrostate $M(0)$, which produces the history of the observable universe, including the macrostate $E$ at the time of observation. }
\end{figure}

Assuming that there are suitable fluctuations in the universe,  we can use them  to explain the origin of the initial low-entropy state described by PH. The universe started in some generic microstate $x_0$ ``chosen at random'' from  phase space (restricted to the energy hypersurface). Most likely it started in  thermal equilibrium, and it will stay in that state for a long time. However, given enough time, it will fluctuate into lower-entropy states. Eventually, it will fluctuate into an extremely low-entropy state---the macrostate $M(0)$ selected by PH. From that state, the universe will grow in entropy, as we have explained on $T_M$. (See Figure 3.) Therefore, the Fluctuation Hypothesis can also explain time's arrow, and it does so without postulating a special initial condition. To summarize, the Fluctuation Theory ($T_F$) consists in the following postulates:
\begin{enumerate}
\item The fundamental dynamical laws. 
\item A uniform probability distribution over all microstates on phase space (restricted to the energy hypersurface). 
\end{enumerate}
Hence, $T_F$ is essentially $T_M$ without PH. To see this, we note that  SP can be understood as the uniform probability distribution that is conditionalized on any initial condition. In $T_M$, SP is conditionalized on PH. In $T_F$, there is no special initial condition, so the probability distribution remains completely uniform over all microstates on  phase space (restricted to the energy hypersurface).

\subsection{Self-Locating Probabilities}

How should we adjudicate between $T_M$ and $T_F$, when both seem to explain time's arrow? One could appeal to super-empirical virtues. For example, $T_F$ seems much simpler than $T_M$: it has fewer axioms.  Moreover, $T_F$ seems less \emph{ad hoc} than $T_M$: in explaining the temporal asymmetry, it does not break temporal symmetry by adding a special initial condition (PH). 

However, things are  more subtle than they seem. There are two lines of reasoning that are often considered. On the first line of reasoning, it seems that $T_F$ is much worse than $T_M$. It is  rare to have thermodynamic fluctuations, and it is extremely rare to have fluctuations that produce an extremely low-entropy state such as $M(0)$. Intuitively, therefore, on $T_F$ it is extremely unlikely to find ourselves living in the current state---a medium-entropy state not too long after the Big Bang and some time away from thermodynamic equilibrium. Most likely, the intuition goes, we should find ourselves in the equilibrium, which is contrary to  our evidence. 

On the second line of reasoning, $T_F$ is no worse than $T_M$. Although it is true that large fluctuations are infrequent, they do occur given enough time. Indeed, large fluctuations will occur with probability close to 1 if we wait long enough. Therefore, in a universe described by $T_F$, some creatures will find themselves in a state that is exactly like our current macrostate. 

Both lines of reasoning seem plausible. The difference is that they are tracking two  kinds of probabilities: \textit{de dicto} probabilities and \textit{de se} (self-locating) probabilities. To appreciate this distinction, we need to understand the distinction between \textit{de dicto} propositions and \textit{de se} (self-locating) propositions. Let us recall Perry’s example (1977) of Lingens who is lost in the Stanford library:

\begin{quotation}
	An amnesiac, Rudolf Lingens, is lost in the Stanford library. He reads a number of things in the library, including a biography of himself, and a detailed account of the library in which he is lost…. He still won’t know who he is, and where he is, no matter how much knowledge he piles up, until that moment when he is ready to say, “\textit{This} place is aisle five, floor six, of Main Library, Stanford. \textit{I} am Rudolf Lingens."\citep[p.492]{PerryFD}
\end{quotation}
In this context, we may understand the sentences “This place is aisle five, floor six, of Main Library, Stanford” and “I am Rudolf Lingens” express self-locating propositions.  Intuitively, we may think of self-locating propositions as only expressible by sentences in which the occurrences of indexical terms such as “This” and “I" are in some sense \textit{essential}, while \textit{de dicto} propositions can be expressed without indexical terms.  On an influential way of thinking (due to \cite{perry1979problem} and \cite{LewisADDDS}), self-locating propositions can have special epistemic significance.\footnote{ Three remarks: (i) Lewis and Perry do not characterize the \textit{de se} / self-location phenomenon in terms of \textit{propositions}. \cite{LewisADDDS} thinks that the previous sentences involving ``This'' and ``I'' express not propositions but property self-ascriptions. \cite{PerryFD} gives another account. For convenience, here I adopt the perspectives and terminology of \cite{egan2006secondary} and \cite{ninan2010se} and take them to express propositions. The difference does not matter to the rest of the arguments. (ii) Not everyone is convinced of the Perry-Lewis view (even after setting aside the issue about property-ascriptions vs. propositions). For some examples of critical perspectives, see \cite{millikan1990myth} and \cite{cappelen2013inessential}. \x{(iii) The main arguments of this paper can still go through even if my arguments don't establish that the relevant propositions and probabilities are distinctively self-locating. If one does not think of them as distinctively self-locating, one can still agree that they are about particular agents, such as you and me. As such, we can agree that they are ``agent-linked'' propositions and probabilities. I can rephrase the summary in \S1: the main arguments in the paper can be seen as a dilemma. Either self-locating probabilities (or agent-linked probabilities, if one is not convinced they are distinctively self-locating) have a place in a scientific} \x{explanation or they do not. Both horns can be problematic. If they do not have a place in a scientific explanation,  the PH-explanation is insufficient. If they do have a place, there is in-principle empirical underdetermination.  }}  With book reading, Lingens the amnesiac can pile up knowledge of  \textit{de dicto} propositions about the locations of books and persons inside the library. However, no matter how much he gains in such knowledge he can still lack knowledge of certain self-locating propositions about who he is and where he is.\footnote{As \cite{ninan2021se} emphasizes,  self-locating (\textit{de se}) propositions can also have behavioral significance, which sets them apart from \textit{de dicto} propositions that happen to be about oneself. Consider Perry's case of bear attack: 
\begin{quotation}
  When you and I both apprehend the thought that I am about to be attacked by a bear, we behave differently. I roll up in a ball, you run to get help. \citep[p.494]{PerryFD}
\end{quotation}
Ninan observes that the distinctive feature of self-locating propositions is their role in the prediction and explanation of action. For example, in certain theoretical frameworks (such as Lewis's centered-world model described below), we can postulate a law-like generalization: necessarily, for all individuals $x$, if $x$ believes \textit{de se} that if she rolls up in a ball, the bear will leave, and $x$ wants the bear to leave, then, if all else is equal, $x'$s having these attitudes will motivate $x$ to roll up in a ball \citep[p.20]{ninan2021se}. } 
Consider another example that is relevant to our discussion of fluctuations: if at time $t$ there are two subjectively indistinguishable copies of me ($A_t$ and $B_t$) in world $w$, even when I am maximally certain about  all the \textit{de dicto} propositions (what the world is objectively like; for example, specified in terms of matter distribution in spacetime), I can still be uncertain about the self-locating proposition that I could express by saying ``\textit{I} am $A_t$.''\footnote{Moreover, believing such propositions can have behavioral significance just as in the bear-attack case. Let us use a case of \cite{elga2004defeating} that is much discussed in the literature. Suppose $B_t$ is Dr. Evil sitting in his impregnable battlestation on the moon and $A_t$ is Dup, the subjectively indistinguishable copy of Dr. Evil created by the Philosophy Defense Force on Earth. If Dup performs actions that correspond to deactivating the battlestation and surrendering, Dup will be treated well; otherwise Dup will be tortured. If Dr. Evil believes in the self-locating proposition that he could express by saying ``\textit{I} am $A_t$,'' assuming he does not want to be tortured, he will be motivated to deactivate the battlestation and surrender. If Dup (or anyone else) believes in that self-locating proposition, assuming he does not want to be tortured, he will (all else being equal) be motivated to act in the same way.}

The earlier characterization in terms of indexicals serves as only an intuitive gloss and need not be taken as a necessary and sufficient condition. It is controversial how to give a precise account of the \textit{de se} / self-location phenomenon. Which model we choose does not impact the rest of the arguments; readers can use their favorite models. For concreteness, I will focus on Lewis's model of centered worlds (1979).\footnote{For some examples of other accounts, see \cite{PerryFD, perry1979problem}, \cite{stalnaker1981indexical, stalnaker2010our}, and \cite{KaplanD}.}  A centered world is a pair of <world, (time-slice of) individual>  \citep[p.147]{LewisADDDS}.  We can think of a possible world as a map of what the world might be like, and a centered world as a map with a ``you are here’’ arrow pointing to a particular individual (Egan 2006, pp.105-106). We may follow Lewis (1979, p.149) and define self-locating probabilities as probability distributions over the space of centered worlds. On this model, \textit{de dicto} probabilities and self-locating (\textit{de se}) probabilities are not that different in structure: both are modeled as probability distributions over some space. The difference lies in the points on that space; the points on the first one being worlds and the points on the second one being centered worlds.\footnote{ Lewis’s definition is a good start, but there are several things worth clarifying. \cite{egan2006secondary} shows that, on the centered-worlds model, propositions \textit{de dicto} can be constructed from self-locating propositions (centered-worlds propositions), making the former a kind of the latter. Egan calls a \textit{de dicto} proposition a \textit{boring} centered-worlds proposition, where a centered-worlds proposition $p$  is \textit{boring} just in case $p$ includes, for each world $w$, either all of the inhabitants in $w$ or none of them. Ninan (2010, p.553) renders the condition as follows:  for any world $w$, and inhabitants $x, y$ in $w$, $<w,x>$  $\in p$ if and only if $<w,y>$  $\in p$. An \textit{interesting} self-locating proposition  is a non-boring centered proposition for which the above condition fails. For our purposes, we can extend Egan's distinction from propositions to probability distributions. A \textit{de dicto} probability distribution is a \textit{boring} kind of self-locating probability distribution such that (1) it is defined over a sample space of centered worlds, and (2) its event space is at least as coarse-grained as  the one constructed   out of the boring self-locating propositions (the \textit{de dicto} ones). We make condition (2) more precise as follows. The  $\sigma$-algebra $\mathscr{F}$ of the boring self-locating probability distribution is one where $\mathscr{F}$ includes at most the following: (a) the entire sample space $\Omega$, the set of all centered worlds, (b) all the boring self-locating propositions, defined above by Egan, and (c) any proposition formed by complementation and countable union from those in (a) and (b). We define an \textit{interesting} self-locating probability distribution to be one whose $\sigma$-algebra includes at least one interesting self-locating proposition. We do not need to further require that the interesting self-locating proposition receives non-zero probability, because if it receives zero probability, then its complement, which is an interesting self-locating proposition, receives non-zero probability. I thank Andy Egan and Isaac Wilhelm for helpful discussions about this issue.  } 
 Intuitively, a \textit{de dicto} probability distribution is defined over only propositions \textit{de dicto}, while a self-locating probability distribution is defined over self-locating propositions (of the interesting kind, in the sense explained in the previous footnote). 
Self-locating probabilities can be used to describe my uncertainty over where I am in space or  time. Recall the earlier example that at time $t$ there are two subjectively indistinguishable copies of me ($A_t$ and $B_t$) in world $w$ such that  I am maximally certain about what the world is objectively like but still uncertain about my self-location (whether I am $A_t$ or $B_t$). This can be represented, say in the case of equal uncertainty of my self-location, by assigning probability $0.5$ to $<w, A_t>$ and probability $0.5$ to $<w,B_t>$.  Self-locating probabilities can count as a kind of agent-linked probabilities,\footnote{ I thank an anonymous referee for suggesting this term.} as the basic points of its probability space are  <world, (time-slice of) individual> pairs, which are made out of not just worlds but also (time-slices of) individual agents.\footnote{Alternatively, we can think of the basic points---the centered worlds---as <world, spacetime point> pairs, where the spacetime point represents a possible location of the individual. This model is suggested (but not endorsed) by \cite{QuinePO}. As Lewis notes, the two models are equivalent in many situations, but having the center as a spacetime point is more restrictive in the sense that we have to assume that no two individuals occupy the same spacetime point. Below, various versions of NPH and the Medium Entropy Hypothesis (MEH) can be understood using either model.}


With the distinction between \textit{de dicto} and \textit{de se} (self-location) in mind, let us go back to the two lines of reasoning. The second line of reasoning tracks \textit{de dicto} probability: the probability that there is a fluctuation in the history of the universe producing the $M(0)$ state is close to 1. In contrast, the first line of reasoning seems to track \textit{de se} probability: the probability that we find ourselves located in such a fluctuation is almost zero, given that fluctuations are so rare and fluctuations of the size of $M(0)$ are rare among all fluctuations. The history of the $T_F$ universe is almost entirely in thermodynamic equilibrium, with most fluctuations being the minimal dips from maximal entropy. But that judgment may implicitly rely on some sort of principle of indifference over self-locating propositions: the probability that we (particular time-slices) are in any particular time-interval is proportional to the length of that interval. For example, the (unconditional) probability that we find ourselves in the first billion years of the universe's history is the same as the probability that we find ourselves  in the second  billion years of the universe's history, and so on. Since the overwhelming majority of times is taken up by thermodynamic equilibrium and not by fluctuations, it is extremely unlikely that we find ourselves in a fluctuation.\footnote{Two remarks: (i) Strictly speaking, on Lewis's centered-worlds model, this may not count as a self-locating probability distribution. If we assume that no individuals can exist at those spacetime points during thermodynamic equilibrium, then there can't be self-locating uncertainty over whether we are at those spacetime points. This is a place where Quine's model of centered worlds in terms of <world, spacetime point> pairs may be more flexible. This principle serves only to introduce more sophisticated principles that do impose  self-locating probability distributions. All the later principles are compatible with Lewis's model and may be thought of as better expressions of the first line of reasoning. 
(ii) This discussion somewhat resembles the debate about fine-tuning argument for design and the multiverse response. The multiverse proponent invokes the \emph{anthropic principle}, emphasizing on the observation-selection effect. The ``this universe'' objection to the multiverse response focuses on the self-locating element in our evidence, which seems to track a different kind of probability. See, for example, \cite{white2000fine} and \cite{manson2003fine} for discussions.   } 

Since much depends on what we mean by  ``our current evidence,'' let me clarify. By that phrase I mean our \emph{direct} evidence, as  we are attempting to examine the justification for our inferential beliefs, including the two hypotheses. Since we are in the context of scientific reasoning, we should be neither too stringent nor too permissive. It seems that we should be realists about the external world. That is, we are justified in believing that we are not brains in vats (BIVs) or Boltzmann Brains. However, it seems inappropriate to take ourselves to have direct access to the exact microstate of our current universe, any states of the past universe, or any states of the future universe. All of them are usually inferred from our direct evidence. For our discussions below, we can entertain different ideas about our evidence that satisfy those constraints. For concreteness, as a first approximation, we stipulate that our current evidence consists in what  Albert (2000) calls the \emph{directly surveyable} condition of the world that it currently happens to be in, i.e. the macrocondition of the world at this instant, which includes, for example, the locations and configurations of galaxies, planets, tables, chairs, observers,  pointers used in detection devices, photographs, and newspapers.\footnote{See Albert 2000, p.96. This is obviously too generous, but it will simplify things. The arguments below are robust with respect to reasonable relaxation of this condition. For the worry that it may be too restrictive, see \S3.6(B). } The central question is whether such evidence is sufficient epistemic ground for either hypothesis. Let us take our current evidence $E$ to be the following:
\begin{description}
\item[Evidence:] Our current evidence $E$ is the medium-entropy macrostate of the universe at this moment. 
\end{description}

How do $T_M$ and $T_F$ compare with respect to our current evidence $E$? Let us adopt a Bayesian framework. Which one has higher \emph{posterior} probability? According to the argument below, it seems that $T_M$ wins the competition:
\begin{tcolorbox}
\centerline{\textbf{The Master Argument}}
\begin{itemize}
\item[P1] Our current evidence $E$ is much less likely on $T_F$  than on $T_M$:  
$$P(E| T_F) \ll P(E| T_M).$$
\item[P2] $T_F$ is  roughly as intrinsically likely as $T_M$: 
$$P(T_F) \approx P(T_M). $$
\item[C]  $T_F$ is much less likely than $T_M$ given our current evidence $E$:  
$$P( T_F|E) \ll P(T_M|E).$$
\end{itemize}
\end{tcolorbox}
The Master Argument is valid by an application of Bayes's theorem (assuming $P(E)\neq 0$):
$$\frac{P( T_F|E)}{  P(T_M|E)} = \frac{P( T_F)}{  P(T_M)} \times \frac{P(E| T_F)}{  P(E|T_M)}.$$


For now, we have adopted a ``uniform'' prior probability over temporally self-locating propositions: we could be anywhere in time, with a probability distribution that is uniform over the entire history (or more or less flat).\footnote{Technically speaking, for a universe without temporal boundaries, the uniform distribution is no longer normalizable. To have a normalizable probability distribution, we could use a Gaussian distribution centered at some point in time. This introduces a temporal bias near the center of the Gaussian. This problem of non-normalizability comes up frequently in cosmology when we consider infinite models. Ideally we would like to find a particular natural choice of measure or probability distribution, which may not exist. For an interesting   philosophical discussion on non-normalizable measures,  responses, and possible pitfalls, see \cite{mcgrew2001probabilities} and the references therein. } Let us call this assumption PoI-De-Se. It is inspired by a restricted version of the principle of indifference. This assumption is simple, but it is arguably an oversimplification. If we do not exist, we cannot observe the universe. Any agent considering what kind of prior probability she should adopt should also take into the \emph{a priori} fact that she exists and has certain conscious experiences. Suppose materialism about the mind is true, then her existence and experiences would require the existence of certain physical states. Suppose  a thinking brain (or something like it) is the minimal requirement for her existence and experience. By taking into account her own existence and experience (which is knowledge \emph{a priori}), she ought to rule out those time intervals containing no brains or no brains that have her experiences. This will produce a  biased probability distribution---she is equally likely to be any brain with her conscious experiences as any other brain. Call this distribution PoI-De-Se*.

Assuming PoI-De-Se*, then, the first premise of the Master Argument does not go through so easily. We are not as likely to be in any interval in time as any other interval. Our temporal location is restricted to those intervals where there are conscious beings with our experiences. However, another problem arises.  The vast majority of fluctuations in the universe are quite small. Typically, they are small deviations from thermal equilibrium. It is very rare for the universe to fluctuate into something like a Big Bang state. It is exponentially more common for it to fluctuate just into the current macrostate. Moreover, it is even more common for it to fluctuate just into a state with no structure at all except for a number of brains thinking the same thoughts we do now (and still more common for it to fluctuate into a state with just one brain). (See Figure 4.) They would have  apparent memories of the past and apparent perceptions of the present (which would be non-veridical). These unfortunate beings are called \emph{Boltzmann brains}. Since there are exponentially more Boltzmann brains than ordinary observers, by PoI-De-Se*, we should be much more confident that we  are Boltzmann brains, which is  absurd. Let us call this  \emph{the Boltzmann brains problem}. 

\begin{figure}[t]
\center
\includegraphics[scale=0.45]{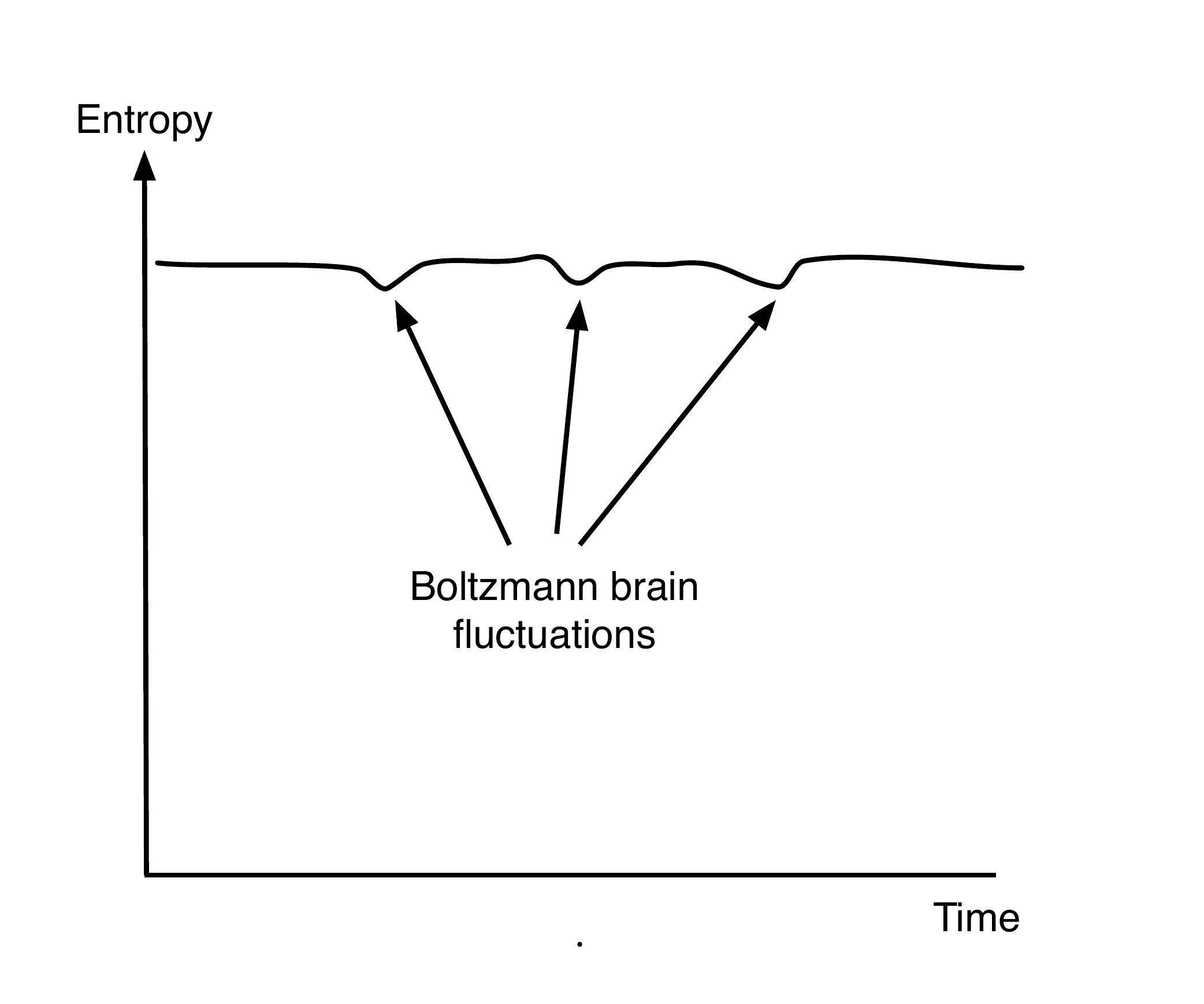}
\caption{Boltzmann Brains}
\end{figure}

In our context, the Boltzmann brains problem comes up as we try to fix $T_F$ by adding a biased self-locating probability distribution. That $T_F$ has this problem seems to be a reason in favor of $T_M$. But is it? Unfortunately, $T_M$ also faces its own version of the Boltzmann brain problem. 

For a universe starting from a low-entropy state (given PH), it will with overwhelming probability (given the Statistical Postulate) increase in entropy. However, after it reaches thermodynamic equilibrium, the highest level of entropy, it will fluctuate downward in entropy if we wait long enough. The smallest such fluctuations compatible with our conscious experiences, which are also the most frequent, will again be the Boltzmann brain fluctuations. They will be the minimal fluctuations compatible with our experiences, ones in which there are a few brains floating temporarily in an environment that is otherwise devoid of any structure. Given $T_M$, it is nonetheless  the case that typical observers (with our experiences) in the universe will be Boltzmann brains. If  PoI-De-Se* is the correct self-locating probability distribution, we are most likely Boltzmann brains, which is absurd.

Here is the upshot of our discussion so far: both $T_M$ and $T_F$ face the Boltzmann brains problem. It is really a problem about temporal self-location. So we will also call it  \emph{the self-location problem}. If I come to believe in the self-locating / centered-worlds proposition that I could express by saying, ``I am the first Boltzmann brain after the universe has first reached thermal equilibrium,'' and if I want to be \textit{accurate} in my beliefs, then I will be motivated to revise many of my beliefs about the past, the present, and the future. For example, I used to believe that I was born to human parents and grew up a normal life, but I now deduce from my new self-locating belief that I was in fact created by a small fluctuation from thermal equilibrium, as the result of a bunch of particles randomly moving together in the shape of a brain having my current thoughts and apparent memories. If I want to be accurate in my beliefs, I will be motivated to revise my belief that I was born to human parents and grew up a normal life.\footnote{That will be an epistemic action similar to the roll-up-in-a-ball action in the bear attack case and the surrender action in the Dr. Evil case. Everyone else sharing those attitudes and the self-locating belief will (all else being equal) also be motivated to perform such an epistemic action and revise their beliefs. See footnotes 18 and 19.} 

To be sure, the problem may seem like just another skeptical hypothesis. Nevertheless, it is worth thinking about what one's response ought to be in this particular case. To try another response: suppose we defend $T_M$ by stipulating that we are not Boltzmann brains, and that we are ordinary observers living in the actual macrostate $E$, which contains not just a few brains but also the normal kind of environment---planet Earth, the solar system, the Milky way, etc.  Suppose further that we have a uniform self-locating probability distribution over (observers living inside) the occurrences of macrostate $E$ (which we assume will occur many times in the long history of the universe). Call this distribution PoI-De-Se**.\footnote{This is related to the indifference principle for self-locating belief \cite{elga2004defeating} advocates for Dr. Evil: similar centered worlds deserve equal credence. Whether they are the same depends on the meaning of Elga's notion of ``similarity.''} This is still not sufficient for getting the monotonic increase of entropy for $T_M$. The minimal fluctuations compatible with $E$ are the medium-entropy dip from thermodynamic equilibrium, in which $E$ is the local minimum of entropy and the entropy is higher in both directions of time. Again, there will be overwhelmingly more minimal fluctuations than  large deviations that first produce  a low-entropy state described by PH and then increase in entropy all the way to $E$. Hence, by the lights of $T_M$ and PoI-De-Se**, it is most likely that we have symmetric histories, just as on $T_F$ and PoI-De-Se**. 

How, then, should defenders of $T_M$ get out of the present conundrum and predict a (typical) monotonic increase of entropy? A possible strategy is to entirely abandon these versions of PoI-De-Se. Instead, they may  choose  a much more biased distribution, called the \textit{Near Past Hypothesis}:
\begin{description}
\item[NPH:] We are currently located in the first epoch of the universe---between the time when PH applies and the first thermodynamic equilibrium; we are equally likely to be any ordinary observers (inside the first epoch) that have our current experiences.\footnote{\cite{winsberg2010, winsberg2012bumps} and \cite{loewer2016earlymentaculus}   postulate a version of NPH for which we are located in the first epoch of the universe. The ideas are similar (and I believe \cite{winsberg2010} gives the principle its name). However, they do not explicitly discuss self-locating probabilities. Their version of  NPH is the non-probabilistic self-locating proposition that we are located in the first epoch of the universe.  I prefer my version for two reasons. First, the non-probabilistic proposition is effectively a self-locating probability distribution that is entirely supported in the first epoch and zero elsewhere. My version of NPH makes explicit that it is a self-locating probability distribution. Second, my version applies to nomologically possible worlds allowed by PH that contain copies of us even in the first epoch. Together with SP, my version of NPH delivers the result that we are most likely normal observers.  In any case, the following arguments (in \S3) apply to both versions. If a defender of $T_M$  insists on using the non-probabilistic version of NPH, they face the same  underdetermination and skeptical problems to be discussed. Those problems are stated for the probabilistic version of NPH but they can be easily restated for the non-probabilistic ones. One just needs to remove the probabilistic parts in the statements of MEH and its variants.} 
\end{description}

Given NPH, our temporal location is restricted to the first epoch of the universe, between  the time when PH applies and the first equilibrium. Thus, we are in the period of ``normal history,'' where fluctuations are probably non-existent. Given our current evidence $E$,  we can predict that  most likely our past had lower entropy and our future will have higher entropy, which meets our goal to predict (with overwhelming probability) a typical increase of entropy around us. NPH is  informative---perhaps too informative for a self-locating distribution. It is intended as an objective norm for the self-locating probability that goes beyond simple indifference.\footnote{There is another place we could postulate such norms---a Many-Worlds interpretation of quantum mechanics in which the Born rule is a self-locating probability not derived from simple axioms but put in by hand. This might be the best strategy forward if none of the existing derivations is compelling.  See \cite{sebens2016self} for an interesting recent attempt of deriving the Born rule from other epistemic principles.} It would be even more compelling if we can derive it from more self-evident principles about rationality. Unfortunately it is hard to see how such a derivation would go. Absent such derivations, we should include NPH as an additional  fundamental postulate in  $T_M$. 

To be sure, it seems odd to have a postulate like NPH in a fundamental theory of physics, even though it plays the same role as explaining the observed temporal asymmetry as PH and SP. The self-locating character of NPH raises many questions. Since it is unlikely to be  derivable from anything else, should we treat it as an objective and fundamental physical law? Alternatively, should we treat it as merely a rationality principle? I am not sure, even though the arguments show that something like NPH is needed.\footnote{For a recent proposal of how to make sense of ``centered'' objective probabilities, see \cite{wilhelm2020centering}.} \cite{winsberg2010} argues that the self-locating character of NPH is sufficient  to reject the fundamentalist vision behind the PH-explanation. Either self-locating probabilities have no place in a scientific theory and should not be invoked in scientific explanations, or they have a place and can be invoked in scientific explanations. For the rest of the paper, I consider the latter. In \S3, I argue that even assuming NPH can appear in a scientific theory, there is a sense in which the explanation cannot be purely scientific because of the existence of robust underdetermination.  

In any case, notwithstanding its self-locating character, NPH gets the work done for $T_M$. Given NPH and our evidence $E$, $T_M$ predicts the correct temporal asymmetry of monotonic entropy growth. Let us use $T_M^{\ast}$ to designate the combined theory of $T_M + NPH$. Hence, we have arrived at the Self-Location Thesis: 

\begin{description}
\item[Self-Location Thesis] The explanation of time's arrow by $T_M$ requires an additional postulate about self-locating probabilities. 
\end{description}

\section{Underdetermination and Skepticism}

In this section, I argue that, if it is permissible to save $T_M$ by adding self-locating probabilities, we may find two different theories, with different implications for time's arrow, that are in-principle underdetermined. That  may open a new door to radical epistemological skepticism. First, I   consider a strategy of strengthening the Fluctuation Theory by using self-locating probabilities. Second, I  introduce ``Boltzmann bubbles.'' Third, I  present a new version of the Master Argument according to which the revised Fluctuation Theory is  on a par with, if not better than, the strengthened Mentaculus Theory $T_M^{\ast}$. Fourth, I  argue that this leads to radical skepticism. Finally, I will consider some possible responses to the skeptical conclusion. 

\subsection{The Medium Entropy Hypothesis}

We saved $T_M$ from the self-location problem by choosing a temporally biased self-locating distribution---NPH. We will now argue from parity and show that we can choose another self-locating distribution to save $T_F$ from the self-location problem.

Recall that $T_F$ has lower posterior probability given $E$ than $T_M$, because $T_F$ assigns low probability to the self-locating proposition $E$ assuming some plausible versions of the principle of indifference.  If we can find a way to make $E$ as probable on $T_F$ as on $T_M^{\ast}$, then $T_F$ would be on a par with $T_M^{\ast}$ with respect to our evidence. In fact, a  strategy exists. Let us add to $T_F$ the following Medium Entropy Hypothesis:
\begin{description}
\item[MEH:] We are currently located in a medium fluctuation of a special kind; we are in a fluctuated state of medium entropy and strong correlations; we are equally likely to be any  observers (inside these states) that have our current experiences.
\end{description}
MEH is similar to NPH,  the self-locating postulate  in $T_M^{\ast}$. Without MEH, and with only some versions of PoI-De-Se, $T_M$ predicts that we are most likely Boltzmann brains, and we are the results of tiny fluctuations. With MEH, our temporal location is restricted to temporal intervals with medium-entropy fluctuations, ones that are much larger deviations from  equilibrium than Boltzmann brain fluctuations (see Figure 5). 
\begin{figure}[t]
\center
\includegraphics[scale=0.45]{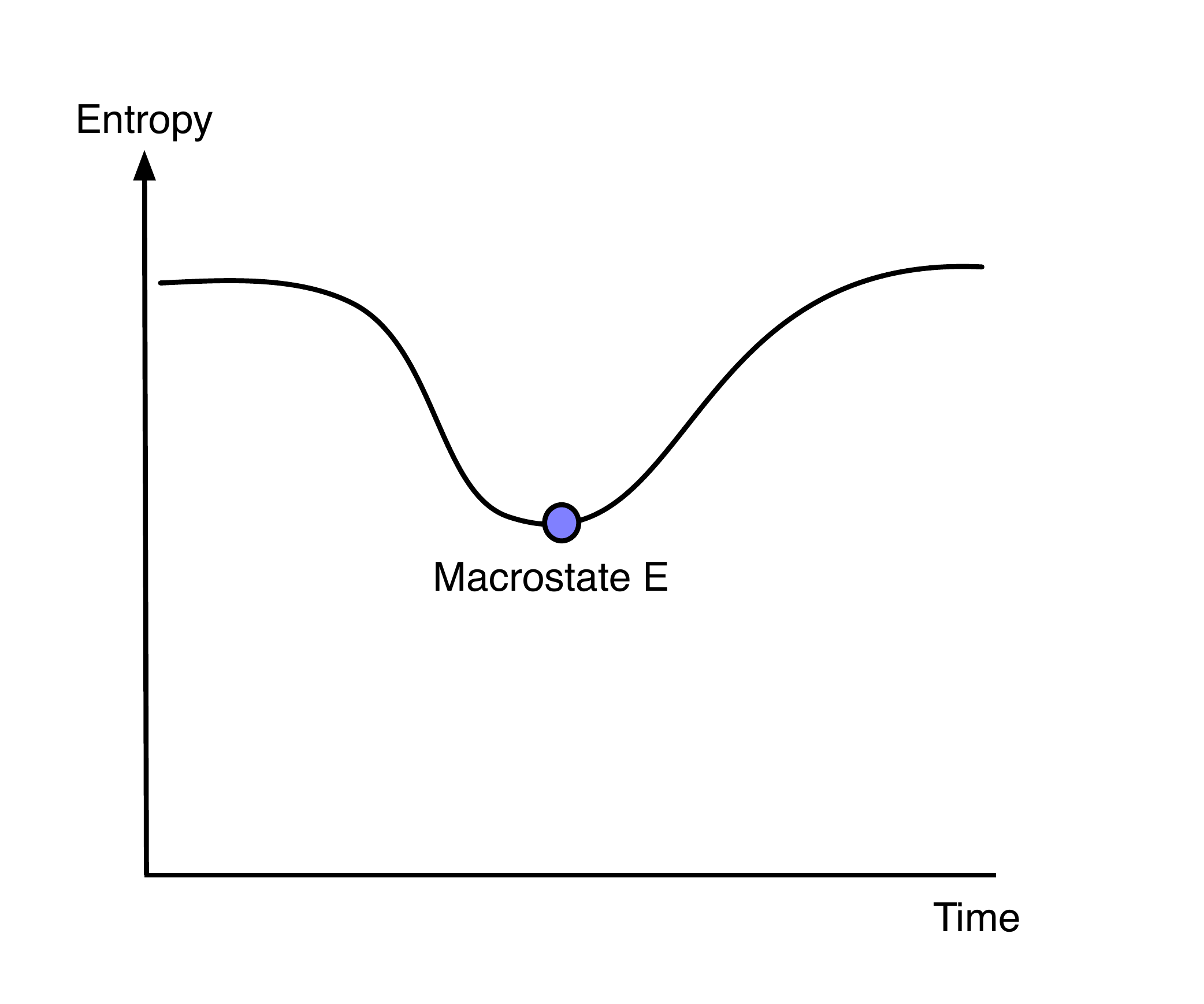}
\caption{Medium Entropy Fluctuations}
\end{figure}
However, they are not as large as the deviations that are required to produce a low-entropy state described by PH. Even so, given that we are located in certain medium-entropy fluctuations, we can infer that most likely we are not Boltzmann brains.\footnote{Given enough time, there will be some medium-entropy fluctuations that are like $E$ but also contain small local fluctuations of Boltzmann brains with our conscious experiences. But these cases are extremely rare. Given the ``uniform'' probability distribution over observers with our conscious experiences, we are very unlikely to be Boltzmann brains. }

(One might reasonably ask: Is MEH  the most natural strategy? Why not just postulate that we are currently located in a period of relaxation following a large fluctuation that resembled the initial state described by PH?  The short answer is that such a large fluctuation is overwhelmingly less frequent to occur than medium fluctuations that just produce the current macrostate. Among versions of the Fluctuation Hypothesis compatible with our current evidence being $E$, other things being equal, it is reasonable to have a prior distribution favoring those versions that postulate we are more likely to be located in shorter and smaller fluctuations than those that postulate we are more likely to be in longer and larger fluctuations. We return to this point in \S3.2 and \S3.5.)

The requirement that the medium fluctuations have to display strong correlations makes the Fluctuation Theory predictively equivalent to the Mentaculus Theory. There are many medium-entropy fluctuations that are abnormal from our point of view, i.e. there are no correlations among different parts of space. 
For example, a medium-entropy fluctuation may contain a photograph of Barack Obama but no real person of Barack Obama; it may contain a left shoe of Napoleon but no right shoe of Napoleon; it may contain a book about pyramids but no real pyramids. These medium-entropy fluctuations, although devoid of the usual correlations among things in space, can nonetheless have medium level of entropy. Their amount of disorder is not lower or higher than the present macrostate $E$. But they are dramatically different from $E$. In fact, most medium-entropy fluctuations are without the right kind of correlations we are used to. That is why we need to add the condition that we are located in the right kind of fluctuations---ones that not only have the right level of entropy but also display strong correlations. 

However, it can be complicated to directly specify what the correlations are. In  $T_M^{\ast}$, PH plays an important role in explaining the correlations. Possible microstates coming out of the PH macrostate will become worlds with the right kind of correlations---they start from a low-entropy state of hot, dense, contracted cosmic soup and will evolve into states with galaxies and more structures. The correlations are already built into the selection of a special low-entropy initial macrostate. We can exploit this feature of PH to define the correlations in the following way:
\begin{description}
\item[Strong Correlations:] The relevant medium-entropy fluctuations are those that  produce macrostates that display the same kind of correlations\footnote{One might object that the ``same kind of correlatioins'' is  vague. However,  any admissible changes to the meaning will not make much of a difference. It is worth remembering that PH as formulated in Boltzmannian statistical mechanics is also a vague postulate. The partition of phase space into macrostates (and the selection of a  particular low-entropy initial macrostate), as introduced in \S2.1 and Figure 1, is  a vague matter. See \cite{chen2018NV} for more discussions.   } as if they evolved from the PH initial condition in the first epoch of the universe. 
\end{description}
That is, the macrostates produced by medium-entropy fluctuations are exactly those macrostates allowed by NPH. Let us call  macrostates allowed by NPH \emph{normal macrostates}. These are the kind of macrostates with strong correlations. Thus, we can adopt the following variant of MEH:
\begin{description}
\item[MEH':] We are currently located in a \emph{normal macrostate} produced by a medium fluctuation; we are equally likely to be any  observers (inside such macrostates) that have our current experiences.
\end{description}
By  locating ourselves in medium fluctuations that produce \emph{normal macrostates}, MEH' provides a probabilistic boost to the  self-locating proposition that $E$ is our current evidence.  In fact, since the possible macrostates are exactly the same on the two theories, the probability  that $T_F + MEH'$  assigns to $E$ is exactly the same as that assigned by $T_M^{\ast}$. Let us use $T_F^{\ast}$ to designate the combined theory of $T_F + MEH'$. 

\subsection{Boltzmann Bubbles}

Before revisiting the Master Argument, let us pause and think where we are and compare the current situation with that of the Boltzmann brains. By the lights of $T_F^{\ast}$, most likely we are in a state of ``local entropy minimum,'' for which entropy is higher in both directions of time. It is produced by a medium-entropy fluctuation of the right sort described by MEH'. As explained before, the fluctuation is not the same kind that produces Boltzmann brains. The current macrostate  has the right sort of structure as we would believe---football stadiums, motorcycles, Jupiter, the Milky way, and etc. Most importantly, there are human beings with physical bodies attached to their brains thinking and living in  normal kinds of environment. The current macrostate is a fluctuation, but it is much larger than the  Boltzmann brain fluctuations.  Thus, the fluctuation  has the right sort of structure  extended in space. We will call it a \emph{Boltzmann bubble fluctuation.} Given just our present evidence $E$, the Boltzmann bubble is an instantaneous macrostate which describes a normal spatial configuration. The minimal fluctuation compatible with a Boltzmann bubble is going to be one in which the Boltzmann bubble is the local entropy minimum, and entropy is higher in both directions of time. 

In contrast to the Boltzmann brain scenarios, the introduction of Boltzmann bubbles makes  $T_F^{\ast}$ much more epistemically conservative. Suppose we have strong philosophical reasons to believe that we are not brains in vats. Then we may have similar strong reasons to believe that we are not Boltzmann brains. (The analogy is not perfect. Unlike BIVs, the existence of Boltzmann brains in the actual world is a salient feature given current physics.) However, our philosophical intuitions  against the possibility that we are in an instantaneous Boltzmann bubble that is extended in space and compatible with our current evidence $E$ are likely weaker and less clear-cut. Even though it may be  implausible that we are Boltzmann brains, it may be less implausible that we are in a Boltzmann bubble. If one worries about the temporal duration of Boltzmann bubbles, we can add that Boltzmann bubbles do not need to be short-lived (unlike typical Boltzmann brains); they can  be extended in time. That is, a medium-level fluctuation may produce a state  that has  lower entropy than  $E$ but develops into $E$ after five days. Can we be in a Boltzmann bubble that is macroscopically indistinguishable  from one coming out of  $T_M^{\ast}$ but only has five days of ``normal'' history? Perhaps we would want a longer ``normal'' history. What about a Boltzmann bubble that has five years of ``normal'' history? By this line of thought, it soon becomes unclear where we draw the line. It is unclear that we can \emph{a priori} rule out (or assign low credence to) the possibilities. Hence, it is unclear that we can \emph{a priori} rule out the possibility that we are in  a Boltzmann bubbles.

One might prefer to live in a Boltzmann bubble that stretches all the way back to a low-entropy state (such as the same initial macrostate described by PH) such that the current macrostate is about 14 billion years away from the entropy minimum of that fluctuation. To live in such a bubble requires a tremendous amount of ``luck.''  Most Boltzmann bubbles compatible with our current evidence \emph{E} are not produced by a large fluctuation. The overwhelming majority of them are in fact produced by the smallest fluctuations that dip  from equilibrium down into \emph{E} and then back up to equilibrium.   Hence, if we compare MEH' with the following hypothesis, it is difficult to (epistemically) justify the assignment of significant credence to it:
\begin{description}
\item[MEH$^{\text{14 billion years}}$:] We are currently located in a \emph{normal macrostate} produced by a large fluctuation whose entropy minimum is about 14 billion years away from us; we are equally likely to be any  observers (inside such macrostates) that have our current experiences.
 \end{description}
 \begin{figure}
\center
\includegraphics[scale=0.24]{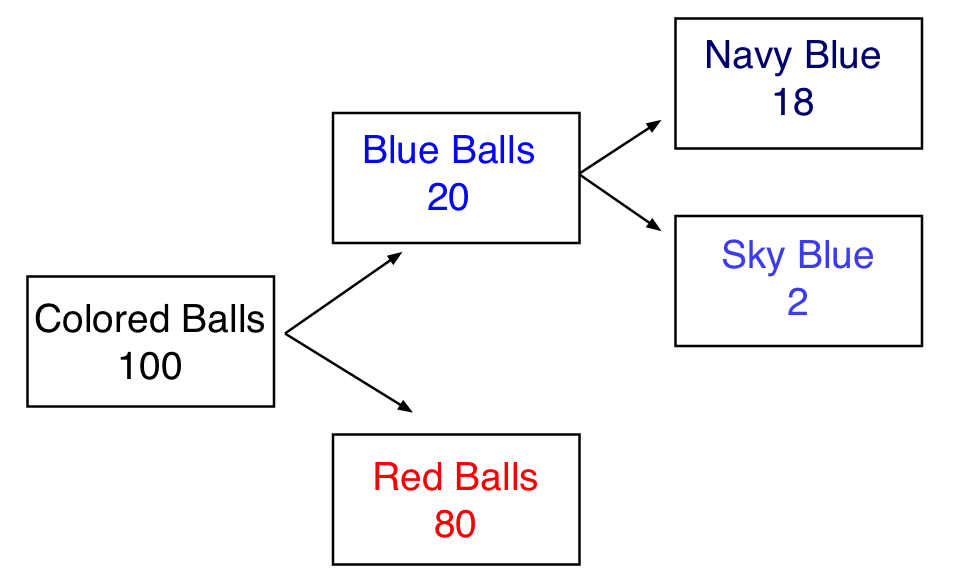}
\caption{Analogy to an urn problem }
\end{figure}
 If MEH' and MEH$^{\text{14 billion years}}$ are both empirically adequate, then the relevant question is how we should assign our priors. Given that there are way more medium-level fluctuations than large fluctuations that dip into a low-entropy macrostate, it seems reasonable to place much higher prior probability in MEH' than in MEH$^{\text{14 billion years}}$.  To illustrate  with an analogy, consider an urn problem drawn in Figure 6. Suppose there are 100 colored balls in the urn and we are going to draw one at random. Suppose for some reason we know that the ball we draw is not red, then it will be a blue ball. Hypothesis 1 says its color is navy blue; Hypothesis 2 says it is sky blue. It seems reasonable to place higher prior probability in the first hypothesis than the second hypothesis. 
 In \S3.5, we return to the question of empirical adequacy of MEH' and see more clearly the relevance of the analogy with the urn problem. For now, we assume that MEH' is compatible (and coheres with) our evidence. Let us thus return to $T_F^{\ast}$, the combined theory of $T_F + MEH'$.

\subsection{The Master Argument$^{\ast}$}

We are ready to show that $T_F^{\ast}$ and $T_M^{\ast}$ are on par with respect to our evidence, which is that the macrostate $E$ is the one we are currently located in. Let us denote the self-locating evidence as $E^{\ast}$. We  can revise the Master Argument as follows:
\begin{tcolorbox}
\centerline{\textbf{The Master Argument$^{\ast}$}}
\begin{itemize}
\item[P1$^{\ast}$] Our current evidence $E^{\ast}$ is  as likely on $T_F^{\ast}$  as on $T_M^{\ast}$:  
$$P(E^{\ast}| T_F^{\ast}) = P(E^{\ast}| T_M^{\ast}).$$
\item[P2$^{\ast}$] $T_F^{\ast}$ is  roughly as intrinsically likely as $T_M^{\ast}$: 
$$P(T_F^{\ast}) \approx P(T_M^{\ast}). $$
\item[C$^{\ast}$]  $T_F^{\ast}$ is roughly as likely as $T_M^{\ast}$ given our current evidence $E^{\ast}$:  
$$P( T_F^{\ast}|E^{\ast}) \approx P(T_M^{\ast}|E^{\ast}).$$
\end{itemize}
\end{tcolorbox}
The Master Argument$^{\ast}$ is valid by an application of Bayes's theorem (assuming $P(E^{\ast})\neq 0$):
$$\frac{P( T_F^{\ast}|E^{\ast})}{  P(T_M^{\ast}|E^{\ast})} = \frac{P( T_F^{\ast})}{  P(T_M^{\ast})} \times \frac{P(E^{\ast}| T_F^{\ast})}{  P(E^{\ast} |T_M^{\ast})}.$$

First, P1$^{\ast}$ follows from the statements of NPH in $T_M^{\ast}$ and MEH' in $T_F^{\ast}$. The only macrostates allowed (to be the current one) by $T_F^{\ast}$ coincides with those allowed by $T_M^{\ast}$. The ratio between macrostates compatible with $E$ and all possible macrostates is exactly the same on the two theories. Hence, the probabilities they assign to our current evidence $E^{\ast}$ are the same. 

Second, there are good reasons to believe that P2$^{\ast}$ is true. I assume  that a simpler and less \emph{ad hoc} theory will be more intrinsically likely than one that is more complex and \emph{ad hoc}.  These are delicate matters of judgment. However, I only argue that the two theories are of the same order of simplicity and \emph{ad hocery}, and that they are roughly as intrinsically likely as each other. On the face of it, MEH' in $T_F^{\ast}$ seems highly \emph{ad hoc}. Given the possibility of so many medium-entropy fluctuations, why choose only the ones compatible with NPH? It seems that we have engineered the result by putting it into the theory by hand. However, a similar question can be asked of NPH in   $T_M^{\ast}$. Given the possibility of locating ourselves in so many different epochs, why choose only the first epoch to be where we can be? So it seems that NPH is equally suspect. Thus, NPH and MEH' may be equally \emph{ad hoc}. Hence, they seem to be tied in this respect.

There is another respect that may be relevant to intrinsic probability. NPH is formulated without reference to MEH', but MEH' is formulated with explicit reference to NPH. The definition of \emph{normal macrostates} invokes NPH. Thus, MEH' may seem more \emph{extrinsic} than NPH, in the sense that it exploits the success of another theory. But it is not clear to me that extrinsicness is a bad thing in this case. If NPH provides a simple way to state the restriction to certain medium-entropy macrostates, then it seems that we can and should use that fact in formulating $T_F^{\ast}$.\footnote{Following \cite{field2016science},  I argue in \citep{chen2017intrinsic} that we should prefer intrinsic postulates to extrinsic postulates in physics, but the arguments are directed at the \textit{de dicto} parts of the physical theories. It is unclear whether they apply to \textit{de se} (self-location) parts of the theory. } In any case, we only need to show that the two theories are roughly equal in intrinsic probability. Even if extrinsicness knocks out some points for $T_F^{\ast}$, as long as the disadvantage is not decisive, the two theories are nonetheless of the same level of intrinsic probability. 

Hence, we have good reasons to accept the conclusion that $T_F^{\ast}$ is roughly as likely as $T_M^{\ast}$ given our current evidence $E^{\ast}$.

\subsection{Skeptical Consequences}

If we accept the conclusion of the Master Argument$^{\ast}$, then we are in trouble. Suppose  that $T_F^{\ast}$  and  $T_M^{\ast}$ are the only two theories currently under consideration. Given the conclusion of the Master Argument$^{\ast}$,  our credence in each theory  should be roughly 0.5.\footnote{To be more realistic, we should consider  more options, for example the possibility that both theories are false. That may well be the case. However, since we do not have any concrete proposal that does better, we should still assign some significant credences to $T_F^{\ast}$  and  $T_M^{\ast}$. To be sure, someone who is convinced of the pessimistic meta-induction argument will not be troubled by the skeptical consequences (or perhaps any conclusion of significance that we draw from contemporary science).  }

As we have explained earlier, while $T_M^{\ast}$ predicts a normal past history, $T_F^{\ast}$ predicts a radically different one. According to $T_F^{\ast}$, most likely we are in a medium-entropy fluctuation. Our future is normal:  entropy will increase and things will appear older. However, our past is very different from what we remembered. In fact, we looked not younger, but older, five years ``ago,'' since the entropy was higher the further we go into the past. The past will not be like a normal past, given that we connect a normal past with the appearance of younger selves. 

$T_F^{\ast}$ predicts a symmetric history, one with entropy growing towards the past and towards the future. This means that our current records and memories about the past are systematically false. My photograph of a five-year old person, with fewer wrinkles and more hair, does not resemble anyone in my past. The person in my past in fact has more wrinkles and fewer hair. In this way, most of our records about the past, which indicate a lower-entropy past, are false. And most of our memories and beliefs about the past are false. The records and memories about the past did not develop from a Big Bang state. Rather, they formed in a random fluctuation out of  thermal equilibrium, creating the \emph{impression} of a low-entropy past. 

If our credence in $T_F^{\ast}$ should be roughly 0.5, then our credence in the following should be roughly 0.5:

\begin{description}
\item[Skepticism about the Past:] Most of our beliefs about the past are false. 
\end{description}

That is not the full extent of the trouble we are in. Many of our beliefs about the present and about the future are partially based on our beliefs about the past (via induction and memory). For example, I believe that tigers are dangerous, based on what I learnt about tigers and biology many years ago. The learning experience probably did not  happen. I believe that the US will approve a COVID-19 vaccine booster shot, based on what I have read in newspapers and online articles. The reading probably did not happen. My memories were likely created by a random fluctuation. 

 If we come to have around 0.5 credence that most of our  beliefs about the past are false, then we should accordingly adjust our credences about the present and the future. Hence, we should significantly lower our credences in  many of our beliefs about the present and the future. We thus enter into a state of agnosticism about many  beliefs that are dear to us. That is a kind of radical epistemological skepticism. The skeptical problem persists even if $T_F^{\ast}$ and $T_M^{\ast}$ are not the only  hypotheses under consideration. If we also consider $T_F$ + MEH$^{\text{14 billion years}}$, the overall probabilistic boost to counter skepticism will be extremely small, since our prior distribution would assign a very low probability on MEH$^{\text{14 billion years}}$ as a large fluctuation that produces a 14-billion year Boltzmann bubble is extremely rare. 

How should we navigate the world if we are convinced by such an argument? I do not yet have an answer. It seems to be a surprising, if not paralyzing, lesson to draw from physics. Physics has challenged many of our cherished beliefs, such as about solidity, space and time, the microscopic reality, and the existence of actual universes outside our own. However, the skeptical consequences we draw from statistical mechanics, and in particular from theories that attempt to explain  the thermodynamic arrow of time, impact some of our core beliefs about the world. Should we find new physical theories that make sure that the skeptical conclusions do not follow? Or should we embrace the surprising consequences as just another conceptual revision required by  physics?  Different people will make different judgments here. 

\cite{carroll2017boltzmann} seems to suggest the first strategy.  In the context of discussions about Boltzmann brains in cosmological theories, he recognizes that the prevalence of Boltzmann brains threatens the epistemic status of the theory. He suggests that we try to find cosmological models in which typical observers are normal people and not Boltzmann brains. The same can be said about Boltzmann bubbles: one could require that cosmological models validate the principle that typical observers do not live in a Boltzmann bubble but have normal history. But again, it is not clear how to define ``normal history'' and it is perhaps vague where the boundary is.  

Perhaps there are other strategies beyond the above two. In the next section, I will discuss  three different strategies.

\subsection{Empirical Incoherence?}

\begin{figure}[t]
\center
\includegraphics[scale=0.45]{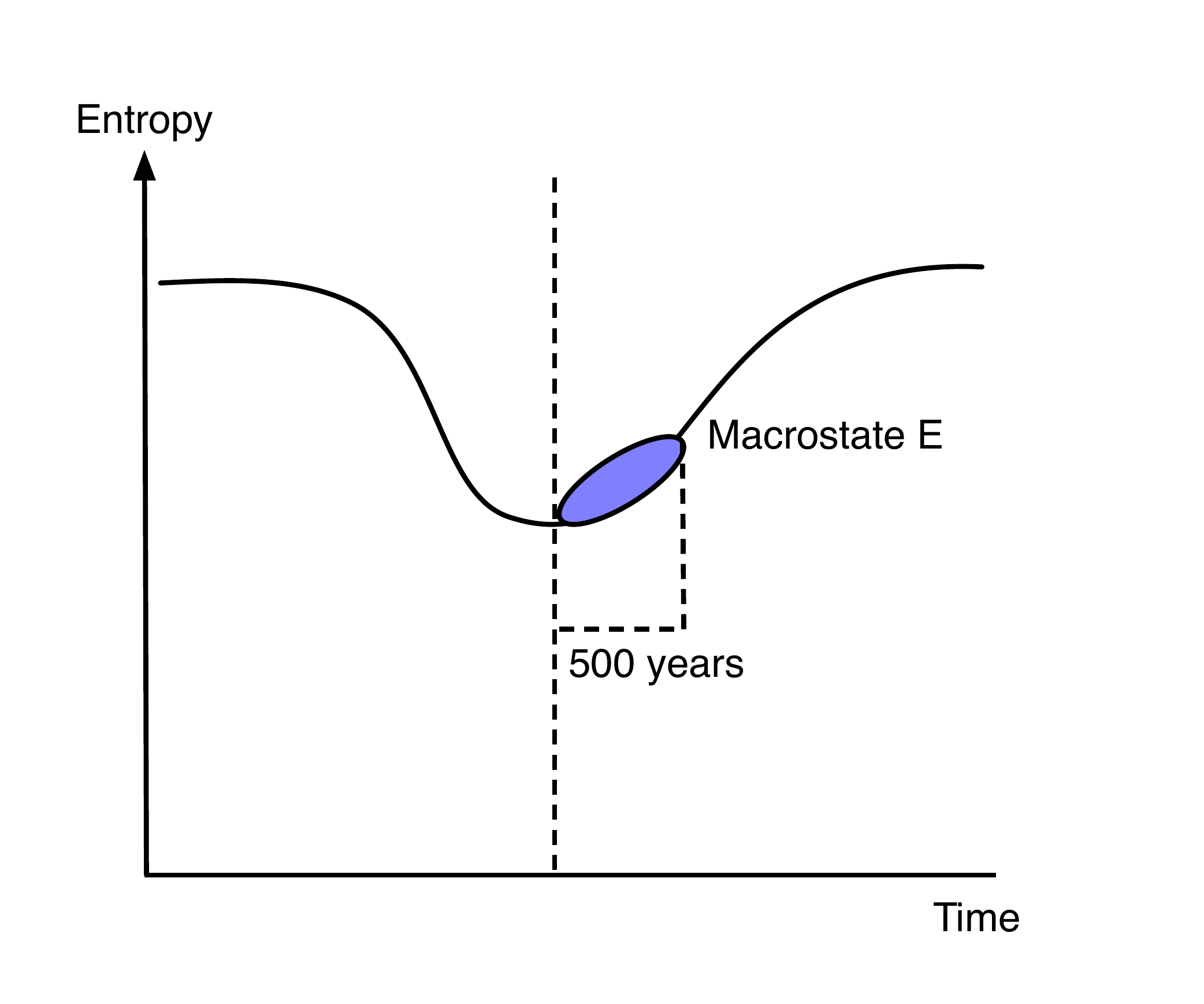}
\caption{A Boltzmann Bubble that extends 500 years.}
\end{figure}

 An initially promising way to respond to the skeptical argument is to point out that $T_F^{\ast}$ is \emph{empirically incoherent}: the theory undermines the empirical evidence we have for accepting it in the first place. This is an \textit{internal} feature any good theory should have. Presumably, we came to consider $T_F^{\ast}$ based on empirical evidence for statistical mechanics and cosmology. However, $T_F^{\ast}$ now predicts that our evidence for accepting it has a significant probability ($\approx 0.5$) to be false. Our memories and records for past experiments and observations are likely the results of random fluctuations from equilibrium. Hence, we have good reasons to reject  $T_F^{\ast}$.\footnote{See \cite{barrett1996empirical, barrett1999quantum} for  discussions about empirical incoherence, especially in the context of quantum theories. In our context of statistical mechanics, the worry is sometimes attributed to Albert (2000) that such theories are ``cognitively unstable.'' See, for example, \cite{carroll2017boltzmann}. However, I am unable to find the reference of cognitive instability in Albert (2000), though he talks about a ``full-blown skeptical catastrophe'' on p.116. } 

However, it is not clear if $T_F^{\ast}$ has to be empirically incoherent.  Suppose the theory is supported by evidence collected in 500 years of normal history. This period includes all the experiments, observations, and derivations we made for classical mechanics, quantum mechanics, cosmology, and statistical mechanics. That is, we  allow $E$ to include not only the current macrostate but also macrostates that stretch to 500 years in the past. Such $E$ can still be produced by medium fluctuations, but we are no longer near the minimum but 500 years away from it. We can revise MEH' as follows:
\begin{description}
\item[MEH$^{\text{500 years}}$:] We are currently located 500 years away from the minimum of a  medium fluctuation that produces a \emph{normal macrostate}; we are equally likely to be any  observers (inside such macrostates) that have our current experiences.
\end{description}
With this version of MEH, $T_F^{\ast}$ predicts that our memories about all the great physics experiments and observations since the scientific revolution were veridical: they really happened. What happened in the middle ages were completely different from what we thought we know, but that does not interfere with the reasons we have for accepting the scientific theory. The Boltzmann bubble which contains us stretches both in space and in time. (See Figure 7.)

In other words, we can let our evidence $E^{\ast}$ to include whatever is necessary to support accepting $T_F^{\ast}$. If that requires 500 years of ``normal'' history, then make the macrostate $E$ stretch back to 500 years. That is still  compatible with accepting $T_F^{\ast}$, which predicts that we most likely live in a Boltzmann bubble and we are 500 years from the minimum of the recent fluctuation.

In fact, MEH$^{\text{500 years}}$  makes $E^{\ast}$ more likely than NPH does, since for each medium fluctuation there are two possibilities for us: either on the ``left'' of the minimum or the ``right.'' This extra probabilistic boost could be significant. For example, it could make even (in posterior probability) any loss to intrinsic probability in $T_F^{\ast}$ after we change MEH' to MEH$^{\text{500 years}}$.

\begin{figure}[t]
\center
\includegraphics[scale=0.25]{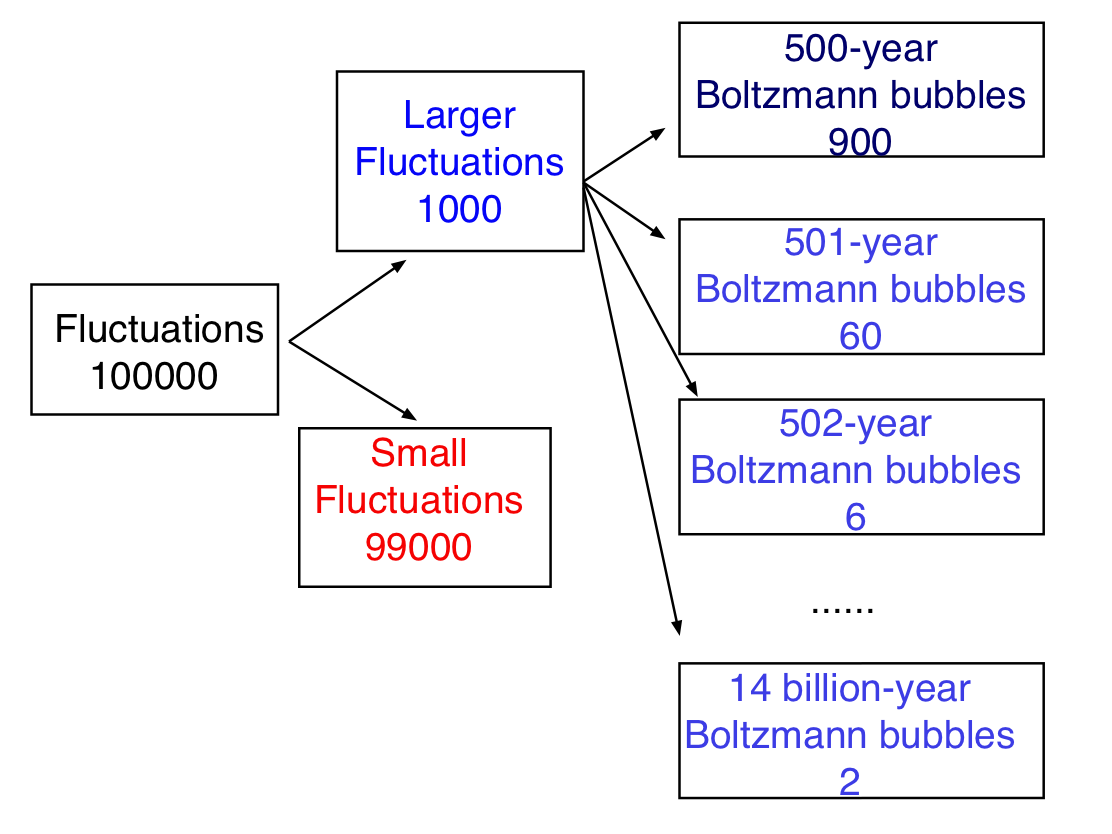}
\caption{A rough comparison of the numbers of different fluctuations (the numbers are illustrative only and do not reflect the relevant scale) }
\end{figure}

But if we are happy to accept MEH$^{\text{500 years}}$ as a modification of $T_F^{\ast}$, why not go further? Why not accept MEH$^{\text{501 years}}$, MEH$^{\text{502 years}}$, ..., or even MEH$^{\text{14 billion years}}$? After all, the longer we stretch the Boltzmann bubble, the more normal history there will be in our specific fluctuation. Other things being equal, wouldn't it be nice to have more normal history? The answer was hinted  in \S3.2. It would indeed be  nice, but I  worry whether that is epistemically the right thing to do. After all, there are overwhelmingly more fluctuations compatible with the selections in MEH$^{\text{500 years}}$ than in MEH$^{\text{501 years}}$. And in fact, it is probably the case that there are overwhelmingly more fluctuations compatible with the selections in MEH$^{\text{500 years}}$ than all the fluctuations compatible with either MEH$^{\text{501 years}}$, MEH$^{\text{502 years}}$, ... , or MEH$^{\text{14 billion years}}$. It seems reasonable to not have higher credence in the disjunction of  MEH$^{\text{501 years}}$, MEH$^{\text{502 years}}$, ... and MEH$^{\text{14 billion years}}$ than in MEH$^{\text{500 years}}$. In Figure 8, we provide a rough comparison of the numbers of different fluctuations. If for \emph{a priori} reasons (such as to avoid  incoherence) we need to postulate our temporal location in larger fluctuations (that are much larger than Boltzmann brain fluctuations and have at least 500 years of normal history), we are still left with some uncertainty over the size of fluctuation we are in. There are much more 500-year Boltzmann bubbles than 501-year Boltzmann bubbles, 502-year Boltzmann bubbles, and 14-billion-year Boltzmann bubbles. Of course, the comparative difference gets exponentially larger as we stretch the Boltzmann bubble from 500 years to 14 billion years (since  fluctuations have to be extremely delicate to dip all the way to something like a Big Bang state). 

By the same reasoning, should we have most of our credence in just MEH' since it is compatible with much more fluctuations than MEH$^{\text{500 years}}$? No, the empirical incoherence of MEH'' is sufficient to subtract much of our credence in that hypothesis. In contrast, we assume that given 500 years of normal history, the fluctuation hypothesis becomes empirically coherent.

\subsection{Other Responses}

\emph{(A) Disputing the Priors.} Another way to respond to the Master Argument$^{\ast}$ is to point out that we should choose priors that overwhelmingly favor $T_M^{\ast}$. This is because if we give some weight to $T_F^{\ast}$, then the previous argument will lead us to skepticism. To lead a successful epistemic life, we ought to choose priors that do not lead us into skepticism. For example, we can assign very low prior probability  to $T_F^{\ast}$ to block P2$^{\ast}$. 

This is what I would like to do in practice. However, is it always epistemically justified? Can we always sweep under the rug any skeptical conclusion we do not like? It would be helpful to have more general principles to guide us here.

In so far as there are any objective norms for credences, I  think that they favor simpler theories. The reason we are warranted to assign extremely low priors to skeptical hypotheses, in many cases, is because the skeptical hypotheses are highly complex.  The Evil Demon Hypothesis,  the Brain-in-Vat Hypothesis, and the Dream Hypothesis are much more complex than the Real World Hypothesis, if we spell them out as detailed theories about the world.\footnote{Take the Evil Demon for example. What laws  describe the demon's actions? Are there laws that together with initial conditions that deterministically or probabilistically determine the course of the demon's actions? Is there a physical model for the deception process? } The Real World Hypothesis can be described with ordinary  simple laws of physics plus the usual initial conditions while the skeptical hypotheses have to be supplemented with extra details that produce the skeptical scenarios. (The Boltzmann Brain Hypothesis may be simpler than $T_F^{\ast}$, but the Boltzmann Brain Hypothesis is not empirically adequate with respect to our current evidence $E^{\ast}$, as we take $E^{\ast}$ to be quite generous and externalistic.) 

In contrast, $T_F^{\ast}$ is on the same level of simplicity as $T_M^{\ast}$: they are similar  cosmological theories with self-locating postulates. If we are warranted to assign low credences in $T_F^{\ast}$, then it must come from other considerations beyond the usual ones based on the complexity of the skeptical theories. However, if such considerations do apply, after careful examination of the case at hand, that would be a welcome result indeed. 

A related response is to criticize $T_F^{\ast}$ for its conspiratorial character.\footnote{I thank Barry Loewer for raising this response and the following response.} Perhaps we should assign low prior probabilities to conspiratorial theories. However, it is hard to give a precise criterion for what makes a physical theory conspiratorial. Here is a proposal: because of MEH', the universe seems to hide its true character from us. It is mostly in thermodynamic equilibrium, and we somehow ended up in the nice temporal regions with the right properties. But that applies equally to $T_M^{\ast}$, so it is unclear why $T_M^{\ast}$ should be higher in prior probability. Superdeterministic theories that attempt to explain Bell-type inequalities in a local way are often called conspiratorial. One reason is that they sometimes demand special initial conditions that are complicated to specify. Our reason for assigning low prior in those theories may just be the general rule  favoring simpler theories, such as Bohmian mechanics and spontaneous collapse theories \citep{ChenBell}.  In our case, as discussed earlier, $T_F^{\ast}$ is not more complicated than $T_M^{\ast}$. 

Another response is to criticize $T_F^{\ast}$ for its ``gruesome'' character. We may endorse a general principle of assigning low prior probabilities to gruesome hypotheses.  If $T_F^{\ast}$ , on pain of being empirically incoherent, requires MEH$^{\text{500 years}}$ instead of MEH', then the theory seems to imply that the past history had an inflection point. Events that happened 500 years before our current time is radically different from current and future events. Hence, MEH$^{\text{500 years}}$ seems to  resemble \cite{GoodmanFFF}'s examples of ``grue'' and ``bleen.'' Perhaps as a general rule we should assign low prior probability to gruesome hypotheses. This is an intriguing response, but I do not know how to develop it further. Presumably, NPH in $T_M^{\ast}$ is also gruesome in the same way. According to NPH and $T_M^{\ast}$, the history of the universe also has inflection points; they are just a bit further away from us in time.

\emph{(B) Disputing the Evidence.} One might object that I have an overly restrictive conception of our current evidence. If we broaden our evidence to include everything that is true and was true, including what the initial time around the Big Bang was like, and how long ago it was from our current time, then the evidential base is significantly enlarged. Call the enlarged body of evidence $E^+$. $E^+$ would  favor $T_M^{\ast}$ over $T_F^{\ast}$. Given the enlarged body of evidence $E^+$, the explanation is entirely scientific. $E^+$ confirms $T_M^{\ast}$ to a much higher degree than $T_F^{\ast}$, and $T_M^{\ast}$ in turn explains the data $E^+$. 

I do not think we should include in our evidence facts extending all the way to the Big Bang. $E^+$   corresponds to (1) a Boltzmann bubble that extends from  macrostate $E$ all the way back to the time of the Big Bang and (2) a self-locating claim that we are currently located in $E$ about 14 billion years from the Big Bang. Our beliefs about the distant past are inferred from scientific theories. If our evidence is already so expansive, and if facts about the distant past are already in our evidence, then we are justified, without consulting any scientific theory, to have beliefs about the actual macroscopic conditions 14 billion years ``ago.'' Now, in this context, we assume that there is no \textit{fundamental arrow of time} that privileges the past or the future. Hence, by symmetry reasoning, we should also be allowed to extend our evidence in the other way: 14 billion years away from now and away from the Big Bang (that is, towards the ``future''). But that is implausible. No one thinks that we are justified to have beliefs about what will actually happen 14 billion years ``in the future'' without consulting a scientific theory. Thus, the objection that relies on such an expansive view of evidence does not work. It seems to rely on a problematic way of reasoning that \cite{price1997cosmology} calls a temporal `double standard.'\footnote{In contrast, the response to the earlier worry (\S3.5) about empirical incoherence is in a different dialectical situation. There,  what leads to that worry is the assumption that our evidence stretches  500 years towards the past. My response is to grant that assumption and show that it can be accommodated by $T_F^{\ast}$.    }

\emph{(C) Fluctuations are impossible in our universe.} Although we have assumed in the paper that fluctuations are physically possible, it is nonetheless relevant to consider the strategy that denies that. If there are no fluctuations, then Boltzmann bubbles cannot form. The Fluctuation Hypothesis would not work. And maybe there are not even Boltzmann brains. In such a universe, Nature is kind to us. Empirical underdetermination would not hold and skeptical consequences would not arise (at least not by my arguments). 

However, how much confidence should we assign to the physical impossibility of fluctuations? It is  unclear as it is still not settled in physics.  Suppose we have reasonable confidence (say, 0.6) in such an impossibility. Then there is still some probability (say, 0.4) that fluctuations can lead to the kind of underdetermination and skeptical worries. In that case, the influence of skeptical hypothesis is much lower (knocking out at most 0.2 credence in many beliefs). However, the influence will still be felt, and a weaker kind of skeptical worry will arise (for many propositions that have threshold credences around 0.5). Thus, it will still be useful to resolve the skeptical worry on the assumption that fluctuations are possible. 

In summary, the above four responses to the skeptical argument are initially promising but ultimately inconclusive. 

\section{Conclusion}

A long-standing problem in the foundations of physics is to explain the arrow of time. A promising explanation points to PH, which is central to the Mentaculus Theory. However, it faces a self-location problem if there are many copies of us in the distant ``past'' or the distant ``future.'' In the first part of this paper, I explained that we need to add self-locating probabilities to save the Mentaculus Theory and derive an increase of entropy in our epoch of the universe. 
In the second part of the paper, I show that such a move leads to underdetermination and opens a new door to skepticism. I use self-locating probabilities to construct a stronger version of the Fluctuation Theory. I argue that our evidence underdetermines between the two theories after we include certain postulates about self-locating probabilities. We are thus led to epistemological skepticism. 

Sadly, the underdetermination is robust: it is not resolved by  appealing to empirical incoherence or simplicity. It is unclear whether there is a decisive and principled reason, besides the desire to avoid skepticism,  for choosing the Mentaculus Theory over the Fluctuation Theory. That is surprising. It seems to me that one of the attractions of the PH research program is the bold vision to eventually provide a scientific explanation, in terms of \textit{objective laws of nature} most confirmed by our current evidence, for our everyday beliefs about the distinction between the past and the future. If the only way to do so is by fiat, by insisting that we need to avoid skepticism and that we must invoke a self-locating probability distribution such as NPH, then it loses that attraction. First, we might doubt whether NPH can be regarded as an objective physical law or part of an objective physical theory. Second, even if NPH can appear in such a scientific explanation, it is unclear why it is superior to its competitor (MEH') in terms of simplicity or adequacy with respect to our current evidence.  Hence, the explanation will fail to be entirely scientific. The original vision faces a difficult dilemma.  

In-principle underdetermination by evidence  occurs elsewhere and is often benign. It is worrisome here for two reasons. First, $T_F^{\ast}$ and $T_M^{\ast}$ are arguably equally simple and invoke the same dynamical laws. That is  not always true in other cases. For example, some might argue that Bohmian mechanics and some version of orthodox (Copenhagen) quantum mechanics are empirically equivalent (insofar as the latter makes determinate predictions), but they are not equally simple and they postulate different dynamical laws. The tie can be broken by invoking  simplicity to favor Bohmian mechanics. Second, the underdetermination here brings forth skepticism and casts doubt on the manifest image of macroscopic arrows of time. That is not always true in other cases, such as the Einsteinian theory of relativity vs. the Lorentzian theory of relativity with a preferred foliation of spacetime \cite[ch.9]{bell2004speakable}. For the latter pair of theories, they both recover the same manifest image of ordinary space and time (although their explanations differ). In our case, they do not recover the same manifest image. $T_F^{\ast}$, although compatible by our current evidence (and equally confirmed as $T_M^{\ast}$), is incompatible with the manifest image of macroscopic temporal asymmetries. 

The worries raised here are at the intersection of epistemology and philosophy of physics. In the end, my conclusions are conditional on the assumptions about what counts as an entirely scientific explanation of time's arrow. I suggest that such an explanation should rely on postulates most confirmed by our current empirical evidence and should not smuggle in assumptions about time's arrow. It is entirely open for someone to resist my conclusion by denying the assumptions. Hence, the worries provide opportunities for defenders of the Mentaculus Theory to clarify and develop the epistemological and theory-choice principles underlying their explanation. Doing so would be valuable contribution to the literature on scientific explanations, underdetermination, and epistemological skepticism.  Moreover, the skeptical problem naturally arises when we reflect on empirical theories and put together a promising explanation of time's arrow with self-locating probabilities. The competing theory is as simple. The problem may be a  manifestation of a more general phenomenon. It would be interesting to think about the connection between this problem and the general question of the nature of self-locating probabilities. Perhaps there is a principled way to think about NPH such that it is the uniquely rational self-locating probabilities to postulate.\footnote{David Albert suggests in personal communication that we may take ``the normal procedures of inference in science'' as an epistemic given. (See also Albert 2000, p.96 and p.129).  Science seems to work pretty well. So we should assume as a given that its methods are reliable. We are rationally obliged to vindicate the normal procedures of inference in science. This may require us to postulate NPH and not MEH' or its variants. } If so, that would also solve the underdetermination and skeptical worries. 

As the case study shows, postulating self-locating probability in physics is like opening a Pandora's box: it is full of conceptual difficulties. We may wonder whether it is appropriate to allow self-locating postulates in physics (probabilistic or not), and if so how to think about their status and how to avoid underdetermination. Reflecting on these questions may teach us something new about the character of physical laws and the nature of scientific explanations.

\section*{Acknowledgement} I thank the editors of \textit{Philosophy and Phenomenological Research} and the anonymous reviewers for valuable comments. I am also grateful to David Albert, Bob Beddor, D Black,  Ben Bronner, Jennifer Carr, Craig Callender, Sean Carroll, David Chalmers, Jonathan Cohen, Antony Eagle, Branden Fitelson, Sheldon Goldstein, Veronica Gomez, Peter Graham, Alan H\'ajek, Christopher Hauser,  John Hawthorne, Christopher Hitchcock,  Michael Huemer, Cameron Kirk-Giannini,     Kelsey  Laity-D'Agostino, Joel Lebowitz,  Barry Loewer,   Tim Maudlin, Bradley Monton,  Oli Odoffin,  Zee Perry, Richard Pettigrew,  Daniel Rubio,   Eric Schwitzgebel, Charles Sebens,   Jonathan Schaffer, Shelly Yiran Shi, Miriam Schoenfield, Katie Steele, Karim Th\'ebault, Michael Tooley,  David Wallace, Isaac Wilhelm,   as well as audiences at  the 2017 AAP Conference, the 2017 Time and Causality in the Sciences Conference (TaCitS), the Sydney University Winter School on Physics and Philosophy of Time, the 3rd Boulder Formal Value Workshop,  CalTech HPS Lecture Series,  Bristol Center for Philosophy of Science, the College of William and Mary, Rutgers Philosophy of Science Group, Rutgers Foundations of Probability Seminar, Rutgers Cognitive Science Grad Talk Series, Wuhan University, UC San Diego philosophy of science reading group, and UC Riverside Workshop on Epistemology and Philosophy of Mind.


\bibliography{test}


\end{document}